\documentclass[11pt,a4paper]{scrartcl}

\usepackage[utf8]{inputenc}

\usepackage{ILD}

\usepackage[symbol]{footmisc}
\usepackage{feynmf}
\usepackage{multirow}
\usepackage{textpos}

\title{Prospects of measuring the branching fraction of the Higgs boson decaying into muon pairs at the International Linear Collider}

\date{09 September 2020}

\addauthor{Shin-ichi Kawada}{\institute{1}}
\addauthor{Jenny List}{\institute{1}}
\addauthor{Mikael Berggren}{\institute{1}}

\addinstitute{1}{DESY, Notkestra{\ss}e 85, 22607 Hamburg, Germany}

\abstract{
The prospects for measuring the branching fraction of $H \to \mu ^+ \mu ^-$ at the International Linear Collider (ILC) have been evaluated based on a full detector simulation of the International Large Detector (ILD) concept, considering centre-of-mass energies ($\sqrt{s}$) of $250$\,GeV and $500$\,GeV.  For both $\sqrt{s}$ cases, the two final states $e^+ e^- \to q\overline{q}H$ and $e^+ e^- \to \nu \overline{\nu}H$ have been analyzed.
For integrated luminosities of 2\,ab$^{-1}$ at $\sqrt{s} =250$\,GeV and 4\,ab$^{-1}$ at $\sqrt{s} =500$\,GeV, the combined precision on the branching fraction of $H \to \mu ^+ \mu ^-$ is estimated to be 17{\%}.
The impact of the transverse momentum resolution for this analysis is also studied\footnote{This work was carried out in the framework of the ILD concept group}.
}

\graphicspath{ {./logos/}{./figures/} }

\begin{document}

\titlepage

\section{Introduction}
\label{sec:intro}
A Standard Model (SM)-like Higgs boson with mass of $\sim 125$~GeV has been discovered by the ATLAS and CMS experiments at the Large Hadron Collider (LHC)~\cite{DiscoveryATLAS,DiscoveryCMS}.
Recently, the decay mode of the Higgs boson to bottom quarks $H \to b\overline{b}$ has been observed at the LHC~\cite{hbbATLAS, hbbCMS}, as well as the $t\overline{t}H$ production process~\cite{tthATLAS, tthCMS}, both being consistent with the SM prediction.
However, there are several important questions to which the SM does not offer an answer: it neither explains the hierarchy problem, nor does it address the nature of dark matter, the origin of cosmic inflation, or the baryon-antibaryon asymmetry in the universe.
Most fundamentally, it does not include gravity.
The Higgs boson could be a portal towards the solution of many of the outstanding questions which are not addressed by the SM.
Thus, it is very important to measure the Higgs boson in as many channels as possible.
In the SM, the Yukawa coupling between matter fermions and the Higgs boson is proportional to the fermion's mass.
If any deviation from this proportionality is observed, it is an indication of new physics beyond the SM.
The size of the deviation from the SM depends on the model, but for a large variety of models, it is estimated to be at the level of a few percent~\cite{deviation}.
To observe such a small deviation, very precise measurements of the properties of the Higgs boson are required.
The International Linear Collider (ILC)~\cite{TDR1, TDR2, TDR31, TDR32, TDR4, LongESU} is one of the future $e^+ e^-$ colliders proposed to deliver this precision with the least possible dependency on models.
The interaction of electrons and positrons will provide a cleaner environment than the proton-proton collisions at the LHC.
Besides, the beams would be longitudinally polarised: $80{\%}$ for electron beam and $30{\%}$ for positron one.
By controlling the polarisation, one can study the chiral structure of the SM interactions and potentially new physics~\cite{EFT}.

In this paper, we focus on the channel of the Higgs boson decays to a pair of muons $H \to \mu ^+\mu ^-$ at the ILC.
This channel is important because it provides an opportunity to measure the Yukawa coupling between the Higgs boson and a second-generation fermion directly.
However, this is a very challenging analysis, because in the SM the branching fraction of $H \to \mu ^+ \mu ^-$ is predicted to be tiny: $2.2 \times 10^{-4}$ for the mass of the Higgs boson of 125~GeV~\cite{HiggsBR}.

At the LHC, the $H \to \mu ^+ \mu ^-$ channel is explored using proton-proton collisions.
In ATLAS, an observed significance of $2.0 \sigma$ with respect to the hypothesis of no $H \to \mu ^+ \mu ^-$ signal was obtained using the full Run 2 dataset of 139~fb$^{-1}$~\cite{ATLASmu_new}.
The CMS observed an excess of events in data with a significance of $3.0 \sigma$~\cite{CMSmu_new}.
The prospects of measuring this channel at the High-Luminosity LHC (HL-LHC) have also been studied.
The ATLAS experiment estimates $\sim 13{\%}$ precision on the signal strength with $3$~ab$^{-1}$ data assuming the phase-II detector upgrade~\cite{ATLASHL-LHC}, based on generator-level samples for the main signal and background processes.
The CMS experiment projects a precision of $\sim 10{\%}$, using a full simulation of an upgraded tracker~\cite{CMSHL-LHC}.
These numbers apply for the signal strength.
On the other hand at the ILC, the measurement of $\sigma \times \mathrm{BF}$ (cross-section times branching fraction) can be turned into a measurement on the BF itself thanks to the total cross-section $\sigma$ being accessible via the so-called recoil technique in a highly model-independent way~\cite{LeptonicRecoil, HadronicRecoil}.
Moreover, by combining other measurements at the ILC, absolute couplings of the Higgs boson can be extracted, \textit{e.g.} based on SM Effective Field Theory~\cite{EFT}.

\begin{figure}[t]
\centering
\includegraphics[width = 7truecm]{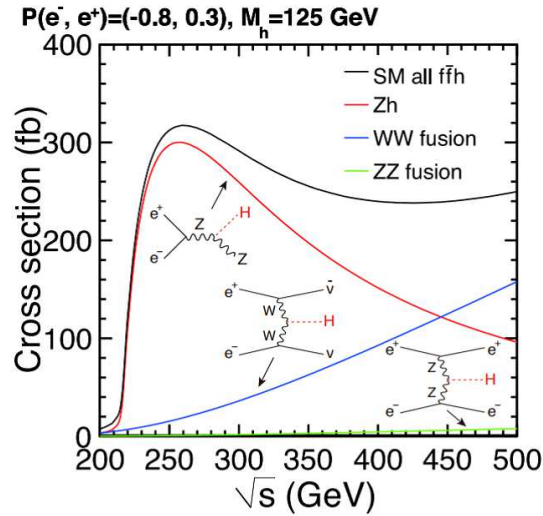}
\caption{The Higgs production cross-section as a function of $\sqrt{s}$.
Taken from Ref.~\cite{ILCPhysicsCase}.}
\label{fig:xsec}
\end{figure}

In this study, the precision expected for the measurement of the branching fraction $\mathrm{BF}(H \to \mu ^+ \mu ^-)$ at the ILC has been estimated based on a full detector simulation of the International Large Detector (ILD) concept~\cite{TDR4, ILDESU} and taking into account all relevant physics and machine-related process.
The Higgs production cross-section as a function of $\sqrt{s}$ at the ILC is shown in Fig.~\ref{fig:xsec}, together with corresponding Feynman diagrams.
The standard running scenario for the ILC has been assumed, which would accumulate 2\,ab$^{-1}$ at $\sqrt{s} =250$\,GeV and 4\,ab$^{-1}$ at $\sqrt{s} =500$\,GeV with beam polarisation sharing as described in Refs.~\cite{ILCOperatingScenario, 250GeVRun}.
Eight different configurations, referred to as analysis channels in the following, have been considered: the two production processes $e^+e^- \to q\overline{q}H \to q\overline{q}\mu ^+ \mu ^-$ and $e^+ e^- \to \nu \overline{\nu}H \to \nu \overline{\nu} \mu ^+ \mu ^-$, with two beam polarisation configurations, and two $\sqrt{s}$ cases.
The case of the electron-positron polarisation combination $\mathcal{P}(e^-,e^+) = (-80{\%}, +30{\%})$ will be referred to as the left-handed case (denoted by L) and $\mathcal{P}(e^-,e^+) = (+80{\%}, -30{\%})$ as the right-handed case (denoted by R).
The cross-section of the $q\overline{q}H$ process changes by about $40{\%}$ between the two beam polarisation configurations, while the $\nu \overline{\nu} H$ process is more significantly affected due to the $WW$-fusion contribution to the $\nu \overline{\nu}$ final state.
At $\sqrt{s} =250$\,GeV, the $e^+ e^- \to ZH$ process is the dominant production process.
Thus, the $ZH \to q\overline{q} \mu ^+ \mu ^-$ channel is the most important signal process at this energy due to the large branching fraction of $Z \to q\overline{q}$.
Since $WW$-fusion is the major production process at $\sqrt{s} =500$\,GeV, $\nu \overline{\nu} H \to \nu \overline{\nu} \mu ^+ \mu ^-$ with left-handed polarisation (including both the $WW$-fusion as well as $ZH$ with $Z\to \nu \bar{\nu}$) is the most relevant channel at this energy.
As the cross sections of the $ZH$ and $WW$-fusion processes will be known to very high precision from other ILC measurements, the separation of the two production modes is not targeted in this paper and it is assumed that the uncertainties on these cross sections will have only a negligible impact on the extraction of the branching fraction.
The expected numbers of $H \to \mu ^+ \mu ^-$ signal events for each channel are summarised in Table~\ref{tab:lumi}, together with the integrated luminosities based on the assumed running scenario.
The fourth column introduces the abbreviations to specify the combination of $\sqrt{s}$, beam polarisation, and signal production process used throughout this paper.
The processes $e^+ e^- \to \ell ^+ \ell ^- H \to \ell ^+ \ell ^- \mu ^+ \mu ^-$, where $\ell$ denotes a lepton ($e$, $\mu$, or $\tau$), are not considered in this paper.

\begin{table}[t]
\centering
\caption{The expected number of signal events $N_{\mathrm{signal}}$ and abbreviations for each channel, where $\int L dt$ is the integrated luminosity based on the running scenario~\cite{ILCOperatingScenario, 250GeVRun}.}
\begin{tabular}{cccccc}
\hline
$\sqrt{s}$ & process & beam pol. & abbreviation & $\int L dt$ (ab$^{-1}$) & $N_{\mathrm{signal}}$ \\
\hline
\multirow{4}{*}{250\,GeV} & \multirow{2}{*}{$q\overline{q}H$} & L & qqH250-L & 0.9 & 41.1 \\
& & R & qqH250-R & 0.9 & 28.1 \\
& \multirow{2}{*}{$\nu \overline{\nu}H$} & L & nnH250-L & 0.9 & 15.0 \\
& & R & nnH250-R & 0.9 & 8.4 \\ \hline
\multirow{4}{*}{500\,GeV} & \multirow{2}{*}{$q\overline{q}H$} & L & qqH500-L & 1.6 & 24.6 \\
& & R & qqH500-R & 1.6 & 16.5 \\
& \multirow{2}{*}{$\nu \overline{\nu}H$} & L & nnH500-L & 1.6 & 57.5 \\
& & R & nnH500-R & 1.6 & 7.9 \\
\hline
\end{tabular}
\label{tab:lumi}
\end{table}

The prospects for measuring the $H \to \mu ^+ \mu ^-$ decay at linear $e^+ e^-$ colliders have been studied previously under various conditions~\cite{mu1, TDR4, mu2, mu3, mu4, mu5}, but all studies except Ref.~\cite{mu5} have been performed at a centre-of-mass energy of $1$\,TeV or higher.
The studies in Refs.~\cite{mu3} and~\cite{mu5} are based on a mass of the Higgs boson of $120$\,GeV.
In Ref.~\cite{mu5} for example, the signal significance is estimated to be $1.1 \sigma$, which corresponds to a precision on $\mathrm{BF}(H \to \mu ^+ \mu ^-)$ of 91{\%}, based on $250$\,fb$^{-1}$ of data at $\sqrt{s} = 250$\,GeV from the analysis of the process $e^+ e^- \to ZH \to q\overline{q}\mu ^+ \mu ^-$, assuming a Higgs mass of $120$\,GeV and the Silicon Detector (SiD) concept~\cite{TDR4, mu5} for the ILC.
Our study comprehensively evaluates the measurement precision of $H \to \mu ^+ \mu ^-$ channel assuming the mass of the Higgs boson of $125$\,GeV and the running scenario of the ILC for $\sqrt{s} =250$\,GeV and $500$\,GeV.

This paper is structured as follows: in Sec.~\ref{sec:ILDMC} the ILD concept and the conditions used for producing the Monte-Carlo (MC) data samples are briefly introduced.
The details of the analysis at $\sqrt{s} =250$\,GeV and 500\,GeV are explained in Sec.~\ref{sec:analysis}.
The impact of the transverse momentum resolution $\sigma_{1/P_t}$ of the central ILD tracking system specifically for this analysis is discussed in Sec.~\ref{sec:impact:ptres} before summarising in Sec.~\ref{sec:summary}.

\section{The ILD Concept and MC Samples}
\label{sec:ILDMC}
ILD~\cite{TDR4, ILDESU} is one of the proposed detector concepts for the ILC.
It is a multi-purpose detector designed for particle flow analysis based on the reconstruction of hadronic jets.
ILD consists of a high precision vertex detector, a time projection chamber, silicon tracking detectors, a highly granular calorimeter system and a forward detector system, all placed inside of a solenoid providing a magnetic field of $3.5$\,T, surrounded by an iron yoke instrumented for muon detection.
Details of the ILD design, as well as about the particle flow concept can be found in Refs.~\cite{TDR4, LongESU, PFA}.

\begin{figure}[t]
\centering
\includegraphics[width = 8truecm]{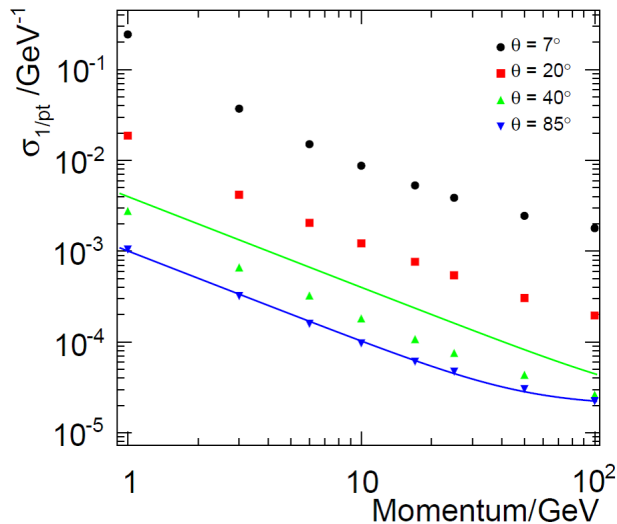}
\caption{The transverse momentum resolution for single muon events as a function of the momentum of particles, for tracks with different polar angles. The points show the resolution as obtained from full simulation of the ILD detector,
the lines correspond to the design goal of $\sigma _{1/P_t} = 2 \times 10^{-5} \oplus 1 \times 10^{-3} / ( P_t \sin \theta )$ for $\theta = 30^{\circ}$ (green) and $\theta = 85^{\circ}$ (blue).
Taken from Ref.~\cite{TDR4}.}
\label{fig:DBD_momres}
\end{figure}

Specifically for this analysis, the key performance aspect of the detector is the transverse momentum resolution $\sigma _{1/P_t}$, since the invariant mass of the muon pair will be the final observable for distinguishing the signal from the background.
The goal of the ILD design for the transverse momentum resolution is $\sigma _{1/P_t} \sim 2 \times 10^{-5} \ \mathrm{GeV} ^{-1}$ at high momenta in the central region of the detector~\cite{goal}.
This level of performance ensures that the model-independent selection of $e^+ e^- \to ZH$ events from the recoil against leptonic $Z \to \mu ^+ \mu ^-$ decays is dominated by beam energy spread rather than by the detector effects~\cite{TDR4}.
This goal is compared to the transverse momentum resolution obtained from the ILD full detector simulation in Fig.~\ref{fig:DBD_momres}. 
The impact of transverse momentum resolution will be discussed in Sec.~\ref{sec:impact:ptres}.
The performance of electromagnetic calorimeter will be important for the recovery of final state radiation photons, which, however, is not yet considered in this study.

The MC samples have been generated in the context of the ILC Technical Design Report~\cite{TDR4} with the matrix element generator \texttt{Whizard}~\cite{Whizard} (version 1.95).
Initial state radiation and the effect of beamstrahlung, as simulated with  \texttt{GuineaPig}~\cite{GuineaPig} based on the beam parameters~\cite{GDE}, are included in the event generation.
\texttt{Pythia}~\cite{Pythia} (version 6.422) is used for parton shower development, hadronisation, and to decay short-lived particles, other than leptons.
The decays of tau leptons are simulated by \texttt{Tauola}~\cite{Tauola1, Tauola2, Tauola3}.
The full detector simulation based on \texttt{Geant4}~\cite{Geant4} has been performed in the \texttt{Mokka} framework~\cite{Mokka} with the so-called \verb|ILD_o1_v05| detector model~\cite{TDR4}.
The pile-up from $\gamma \gamma \to$ low $P_t$ hadron events has been generated based on the cross-section model described in Ref.~\cite{Overlay}.
These events have been passed through the same \texttt{Geant4}-based detector simulation and the resulting hits were overlaid to all MC samples before the reconstruction.
Events have been reconstructed using \texttt{PandoraPFA}~\cite{PFA} in the \texttt{Marlin} framework~\cite{Marlin}.

To make the analysis as realistic as possible, all relevant SM processes with up to six fermions in the final state have been included.
For the ILC-TDR~\cite{TDR4}, the SM background samples are grouped with the number of fermions in the final state.
For example, the  $e^+ e^- \to 2f$ process comprises the SM processes with two fermions\footnote{In this context, fermions and anti-fermions are counted as fermions.} in the final state, \textit{i.e.}, two quarks or two leptons.
Table~\ref{tab:ListofMC} shows the list of MC samples used in this analysis.
The total number of simulated events are of the order of $10^7$ for each centre-of-mass energy.
For the production of these MC samples, a significant amount of CPU time was necessary.
A new MC production of similar size in the context of the recently completed ILD Interim Design Report~\cite{ILDIDR} required about 320 CPU-years.

\begin{table}[t]
\centering
\caption{List of MC samples used in this analysis.}
\begin{tabular}{c|cc}
\hline
 & $\sqrt{s} =250$ GeV & $\sqrt{s} =500$ GeV \\
\hline
Processes & $e^+ e^- \to 2f$ & $e^+ e^- \to 2f$ \\
 & $e^+ e^- \to 4f$ & $e^+ e^- \to 4f$ \\
 & & $e^+ e^- \to 6f$ \\
 & $e^{\pm} \gamma \to 3f$ & $e^{\pm} \gamma \to 5f$ \\
 & $\gamma \gamma \to 2f$ & $\gamma \gamma \to 4f$ \\
 & $e^+ e^- \to f\overline{f}H$ & $e^+ e^- \to f\overline{f}H$ \\
\hline
\end{tabular}
\label{tab:ListofMC}
\end{table}

In the $q\overline{q}H$ analysis, the process $e^+ e^- \to q\overline{q}H$ with $H \to \mu ^+ \mu ^-$ is considered as the signal, and all other processes, including other Higgs channels, are considered as background.
Similarly, in the $\nu \overline{\nu} H$ analysis, $e^+ e^- \to \nu \overline{\nu} H \to \nu \overline{\nu} \mu ^+ \mu ^-$ is considered as signal, while all other processes are regarded as background.

\section{Analysis}
\label{sec:analysis}
The analysis is structured in the same way for all channels: first, a pair of well-measured, prompt, oppositely charged muons consistent with $H \to \mu ^+ \mu ^-$ is selected by a series of sequential cuts described in Sec~\ref{subsec:analysis:muid}.
These cuts are ``common cuts'' for all analysis channels since they only pertain to the properties of the $H \to \mu ^+ \mu ^-$ signal-candidates.
Then, the rest of the event is subjected to channel-specific event reconstruction and event selection as detailed in Sec.~\ref{subsec:analysis:isr} to~\ref{subsec:analysis:presel}.
To perform the final event selection, a multivariate analysis technique is used, as described in Sec.~\ref{subsec:analysis:mva}.
Finally, $\mathrm{BF} (H \to \mu ^+ \mu ^-)$ is extracted from a template fit to the invariant di-muon mass distributions in each channel.
A toy MC technique is applied to estimate the final precision.
This technique and its results will be discussed in Sec.~\ref{subsec:analysis:toymc} and~\ref{subsec:analysis:result}.

\subsection{Identification of $H \to \mu ^+ \mu ^-$ Candidates}
\label{subsec:analysis:muid}
First, the \texttt{IsolatedLeptonTagging} processor~\cite{Tagging} is applied to select the $H \to \mu ^+ \mu ^-$ candidates.
The criteria required for isolated muon candidates are listed in Table~\ref{tab:muonselection}.
Here, $E_{\mathrm{CAL}}$ is the total energy deposit in the calorimeter system (apart from the BeamCal), $p_{\mathrm{track}}$ is the momentum of the track, $E_{\mathrm{yoke}}$ is the energy deposit in the yoke, $d_0$ and $z_0$ are the impact parameters in transverse and longitudinal directions~\cite{LCIOTrack}, respectively, with their uncertainties $\sigma (d_0)$ and $\sigma (z_0)$ as obtained for each individual track fit.
A multivariate double-cone method is used to check the isolation of each particle, and a cut on the MVA output is applied.
In most cases, the default values of the \texttt{IsolatedLeptonTagging} processor are used for isolated muon identification.
In the case of the nnh250-L/R channels, the signal rate is rather low to start with, while the events hardly contain any other particles than the muons.
Therefore, some of the criteria have been relaxed for these channels.
The particles passed all the requirements listed in Table~\ref{tab:muonselection} are considered as isolated muon candidates.
Only events that have exactly one $\mu ^+$ and one $\mu ^-$ are considered for further analysis.

\begin{table}[t]
\centering
\caption{Requirements for selecting $H \to \mu ^+ \mu ^-$ candidate in the \texttt{IsolatedLeptonTagging} processor.
The definition of variables is in the text.}
\begin{tabular}{c|cccc}
\hline
variables & qqH250-L/R & nnH250-L/R & qqH500-L/R & nnH500-L/R \\
\hline
$E_{\mathrm{CAL}} / p_{\mathrm{track}}$ & $< 0.5$ & $< 0.3$ & \multicolumn{2}{c}{$< 0.5$}  \\
$p_{\mathrm{track}}$ (GeV) & \multicolumn{2}{c}{$> 5$}  & \multicolumn{2}{c}{$> 10$}  \\
$E_{\mathrm{yoke}}$ (GeV) & \multicolumn{4}{c}{$> 0.5$}  \\
$|d_0 / \sigma (d_0)|$ &  \multicolumn{4}{c}{$<5$} \\
$|z_0 / \sigma (z_0)|$ &  \multicolumn{4}{c}{$<5$}  \\
MVA cut & \multicolumn{4}{c}{$> 0.7$} \\
\hline
\end{tabular}
\label{tab:muonselection}
\end{table}

\begin{figure}[t]
\centering
\includegraphics[width = 10truecm]{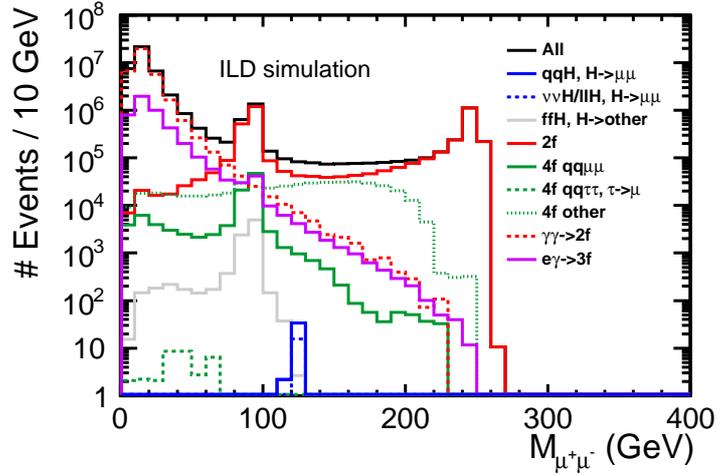}
\caption{The distribution of $M_{\mu ^+ \mu ^-}$ before applying cut {\#}5 (see Table~\ref{tab:hmumucuts}) in qqH250-L.
The solid blue histogram shows the signal process.}
\label{fig:mumu_mass_qqh250L}
\end{figure}

The ``common cuts'' applied to the $H \to \mu ^+ \mu ^-$ candidates are summarised in Table~\ref{tab:hmumucuts}.
They have been chosen by maximising efficiency times purity.
Here, $\chi ^2/\mathrm{Ndf}(\mu^{\pm})$ is the reduced $\chi ^2$ of the muon track fit, $\sigma (M_{\mu ^+ \mu ^-})$ is the event-by-event mass uncertainty as obtained from error propagation from the track parameter uncertainties, $M_{\mu ^+ \mu ^-}$ is the invariant mass of $H \to \mu ^+ \mu ^-$ candidate, and $\theta _{\mu ^+ \mu ^-}$ is the angle between the $\mu ^+$ and the $\mu ^-$.
Cut {\#}1 ($\chi ^2 / \mathrm{Ndf}$) serves to select well-measured tracks, followed by two cuts ($d_0$ and $z_0$) which ensure prompt tracks and reject muons likely to originate from $\tau$ decays.
The cut on $\sigma (M_{\mu ^+ \mu ^-})$, cut {\#}4, rejects events with too imprecise mass measurement, while cut {\#}5 on $M_{\mu ^+ \mu ^-}$ ensures that the invariant mass of the candidate is well above $M_Z$, while not removing any di-muons with a mass close to $M_H$.
Fig.~\ref{fig:mumu_mass_qqh250L} shows the distribution of $M_{\mu ^+ \mu ^-}$ before applying cut {\#}5 for the example of the qqh250-L channel.
The last cut ({\#}6, $\cos \theta _{\mu ^+ \mu ^-}$) requires the di-muons to have a minimum opening angle which is defined by the boost of the produced Higgs boson and thus depends significantly on the centre-of-mass energy.

\begin{table}[t]
\centering
\caption{The ``common cuts'' for $H \to \mu ^+ \mu ^-$ candidate.
The definition of variables is in the text.}
\begin{tabular}{c|c|cccc}
\hline
\# & variables & qqH250-L/R & nnH250-L/R & qqH500-L/R & nnH500-L/R \\
\hline
1 & $\chi ^2/\mathrm{Ndf}(\mu ^{\pm})$ & \multicolumn{4}{c}{$0.5 - 1.5$}\\
2 & $|d_0(\mu ^{\pm})|$ (mm) & \multicolumn{4}{c}{$<0.01$} \\
3 & $|z_0(\mu ^-) - z_0(\mu ^+)|$ (mm) & \multicolumn{4}{c}{$<0.5$} \\
4 & $\sigma (M_{\mu ^+ \mu ^-})$ (GeV) & \multicolumn{2}{c}{$<0.5$} & \multicolumn{2}{c}{$<1$} \\
5 & $M_{\mu ^+ \mu ^-}$ (GeV) & \multicolumn{4}{c}{$100 - 130$} \\
6 & $\cos \theta _{\mu ^+ \mu ^-}$ & \multicolumn{2}{c}{$<-0.45$} & \multicolumn{2}{c}{$<0.55$}\\
\hline
\end{tabular}
\label{tab:hmumucuts}
\end{table}

\subsection{ISR Identification}
\label{subsec:analysis:isr}
Some events include energetic initial state radiation (ISR) photons which, if within the detector acceptance, will affect the further analysis.
Therefore a simple ISR identification procedure is applied after the muon identification.
First, a candidate photon is selected if its energy $E_{\mathrm{photon}}$ is greater than 10\,GeV.
All charged particle energies in a cone with half-opening angle $\cos \theta _{\mathrm{cone}} = 0.95$ around the photon are summed up.
If this energy sum is less than 5{\%} of the photon energy, the photon is regarded as an ISR photon.
These ISR photons are not subject to jet reconstruction.

\subsection{Jet Reconstruction}
\label{subsec:analysis:jets}
For the $q\overline{q}H$ channels, a jet clustering algorithm is applied to reconstruct $Z \to q\overline{q}$ candidates.
After the selection of $H \to \mu ^+ \mu ^-$ and a possible ISR photon, one can expect that the remaining particles consist of $Z \to q\overline{q}$ and some contribution from the overlaid $\gamma \gamma \to$ low $P_t$ events.
At $\sqrt{s} = 250$\,GeV, only $0.4$ $\gamma \gamma \to$ low $P_t$ hadron events are expected per bunch crossing on average~\cite{Overlay}.
Thus, no dedicated attempt is made to remove the overlay and the Durham clustering algorithm~\cite{Durham} is used to force the remaining particles into two jets.
However at $\sqrt{s} = 500$\,GeV, the average number of $\gamma \gamma \to$ low $P_t$ hadron events per bunch crossing increases to $1.7$~\cite{Overlay}.
To remove this background, an exclusive $k_T$ clustering algorithm~\cite{kT1, kT2} is applied to the remaining particles, requesting four jets with a generalised jet radius of 1.0. 
The jet radius has been tuned to optimise the reconstruction of the invariant mass spectrum of the $Z \to q\overline{q}$ system.
The clustering into four jets has been proven to render the overlay-removal step more robust in the presence of hard gluon emission~\cite{ThesisMadalina}.
After this process, the Durham algorithm~\cite{Durham} is used to force the particles contained in the four $k_T$-jets into two final jets.
The $H \to \mu ^+ \mu ^-$ candidates and ISR photons are not included in jet clustering.

\subsection{Preselection}
\label{subsec:analysis:presel}
After the general preselection described in Sec.~\ref{subsec:analysis:muid}, channel-specific cuts are applied.
Table~\ref{tab:qqhcuts} summarises the cuts applied in the $q\overline{q}H$ channels.
Cut {\#}1 vetoes isolated leptons since no further isolated leptons beyond the $H \to \mu ^+ \mu ^-$ candidate are expected in the event.
For this cut, the \texttt{IsolatedLeptonTagging} processor~\cite{Tagging} is applied again to the remaining particles and requires that no isolated leptons (electrons or muons) are found.
The cut on the number of charged particles in each jet ({\#}3) is applied to remove reconstructed jets of low charged multiplicity originating mostly from 1-prong hadronic tau decays.
The last cut {\#}4 selects events with the invariant mass $M_{jj}$ of two jets consistent with $Z \to q\overline{q}$.
Since the di-jet mass resolution is somewhat worse at $\sqrt{s}=500$\,GeV, the allowed mass window is wider than in the $\sqrt{s}=250$\,GeV case.

\begin{table}[t]
\centering
\caption{The channel-specific cuts for the $q\overline{q}H$ processes.
The definition of variables is in the text.}
\begin{tabular}{c|c|cc}
\hline
\# & quantity & qqH250-L/R & qqH500-L/R \\
\hline
1 & number of isolated leptons beyond $\mu ^+ \mu ^-$ pair & \multicolumn{2}{c}{$0$} \\
2 & jet clustering successful &  \multicolumn{2}{c}{yes}\\
3 & number of charged particles in each jet & \multicolumn{2}{c}{$\ge 2$} \\
4 & $M_{jj}$ (GeV) & 50 - 130 & 50 - 160 \\
\hline
\end{tabular}
\label{tab:qqhcuts}
\end{table}

The channel-specific cuts for the $\nu \overline{\nu} H$ channel are summarised in Table~\ref{tab:nnhcuts}.
Apart from the $H \to \mu ^+ \mu ^-$ candidate, there should be no visible particles in the event except for some contribution from $\gamma \gamma \to$ low $P_t$ hadron backgrounds which typically have low transverse momentum.
Therefore, the number of charged particles in an event, excluding the $H \to \mu ^+ \mu ^-$ candidate, which have a transverse momentum larger than 5\,GeV,  $N_{P_t}$, has to be zero (cut {\#1}).
The four-momenta of all particle flow objects in an event are summed up to the visible four-momentum, of which $E_\mathrm{vis}$ and $P_t$ are the energy and the transverse momentum, respectively.
The visible four-momentum can be subtracted from the four-momentum of the initial state, $(\sqrt{s}, \sqrt{s} \cdot \tan \theta _{\mathrm{cross}} / 2, 0, 0)$ where $\theta _{\mathrm{cross}} = 14~\mathrm{mrad}$ is the crossing angle of beam collision, to obtain the missing four-momentum, and in particular its polar angle, $\theta _{\mathrm{miss}}$.
The cuts {\#}2 to {\#}4 ($E_{\mathrm{vis}}$, missing $P_t$, and $\cos \theta _{\mathrm{miss}}$) use these quantities to select events with neutrinos.
The $E_{\mathrm{vis}}$ requirement thereby depends on the centre-of-mass energy.

\begin{table}[t]
\centering
\caption{The channel-specific cuts for $\nu \overline{\nu}H$ processes.
The definition of variables is in the text.}
\begin{tabular}{c|c|cc}
\hline
\# & variables & nnH250-L/R & nnH500-L/R \\
\hline
1 & $N_{P_t}$ & \multicolumn{2}{c}{$0$} \\
2 & $E_{\mathrm{vis}}$ (GeV) & 120 - 170 & 120 - 350 \\
3 & missing $P_t$ (GeV) &  \multicolumn{2}{c}{$>5$}\\
4 & $|\cos \theta _{\mathrm{miss}}|$ &  \multicolumn{2}{c}{$<0.99$} \\
\hline
\end{tabular}
\label{tab:nnhcuts}
\end{table}

Table~\ref{tab:cuttable} shows the number of signal and background events in each channel after the preselection.
Overall, the signal selection efficiency is $\sim 85{\%}$ in all channels.
The ``other Higgs'' category includes all events with Higgs bosons with other decay modes than the signal.
The category of ``irreducible'' backgrounds is defined as follows: for the $q\overline{q}H$ process, $e^+ e^- \to 4f \to qq\mu ^+ \mu ^-$ and $qq \tau ^+ \tau ^-$ with both tau leptons decaying into $\mu$ are defined as irreducible.
For the $\nu \overline{\nu} H$ process, $e^+ e^- \to 4f \to \nu \nu \mu ^+ \mu ^-$, $\nu \nu \mu \tau$ with $\tau$ decaying into $\mu$, and $\nu \nu \tau ^+ \tau ^-$ with both $\tau$ decaying into $\mu$ are defined as irreducible.
These events have the same or very similar final states to the signal, thus they are difficult to remove.
After the preselection, the irreducible category dominates the background in nearly all analysis channels.  In case of the qqH500-L/R channels, though, the $e^+ e^- \to 6f$ processes (dominated by $e^+ e^- \to t\overline{t}$) remain at the same level as the irreducible background.

\begin{table}[t]
\centering
\caption{The number of signal and background events in each channel after the channel-specific cuts, weighted to the beam polarisation and luminosity settings given in Table~\ref{tab:lumi}.
The definition of ``irreducible'' is in the text.
Numbers in brackets show the signal selection efficiency.}
\begin{tabular}{c|cccc}
\hline
\multirow{2}{*}{channel} & \multirow{2}{*}{signal} & other & \multirow{2}{*}{``irreducible''} & other SM \\
 & & Higgs & & background \\
\hline
qqH250-L & 36 (88{\%}) & 143 & $3.54 \times 10^3$ & 335  \\
qqH250-R & 25 (87{\%}) & 106 & $1.68 \times 10^3$ & 123 \\
nnH250-L & 13 (87{\%}) & 1.1 & $4.88 \times 10^4$ & 670 \\
nnH250-R & 7.3 (88{\%}) & 1.1 & $3.92 \times 10^3$ & 581 \\
qqH500-L & 20 (82{\%}) & 99.9 & $2.53 \times 10^3$ & $2.15 \times 10^3$ \\
qqH500-R & 14 (82{\%}) & 64.5 & $1.15 \times 10^3$ & $1.18 \times 10^3$ \\
nnH500-L & 49 (86{\%}) & 4.4 & $9.71 \times 10^3$ & 933 \\
nnH500-R & 6.7 (85{\%}) & 0.6 & $1.09 \times 10^3$ & 595 \\
\hline
\end{tabular}
\label{tab:cuttable}
\end{table}


\subsection{Multivariate Analysis}
\label{subsec:analysis:mva}
\begin{figure}[htbp]
\begin{center}
\begin{subfigure}{0.45\textwidth}
  \includegraphics[width = \textwidth]{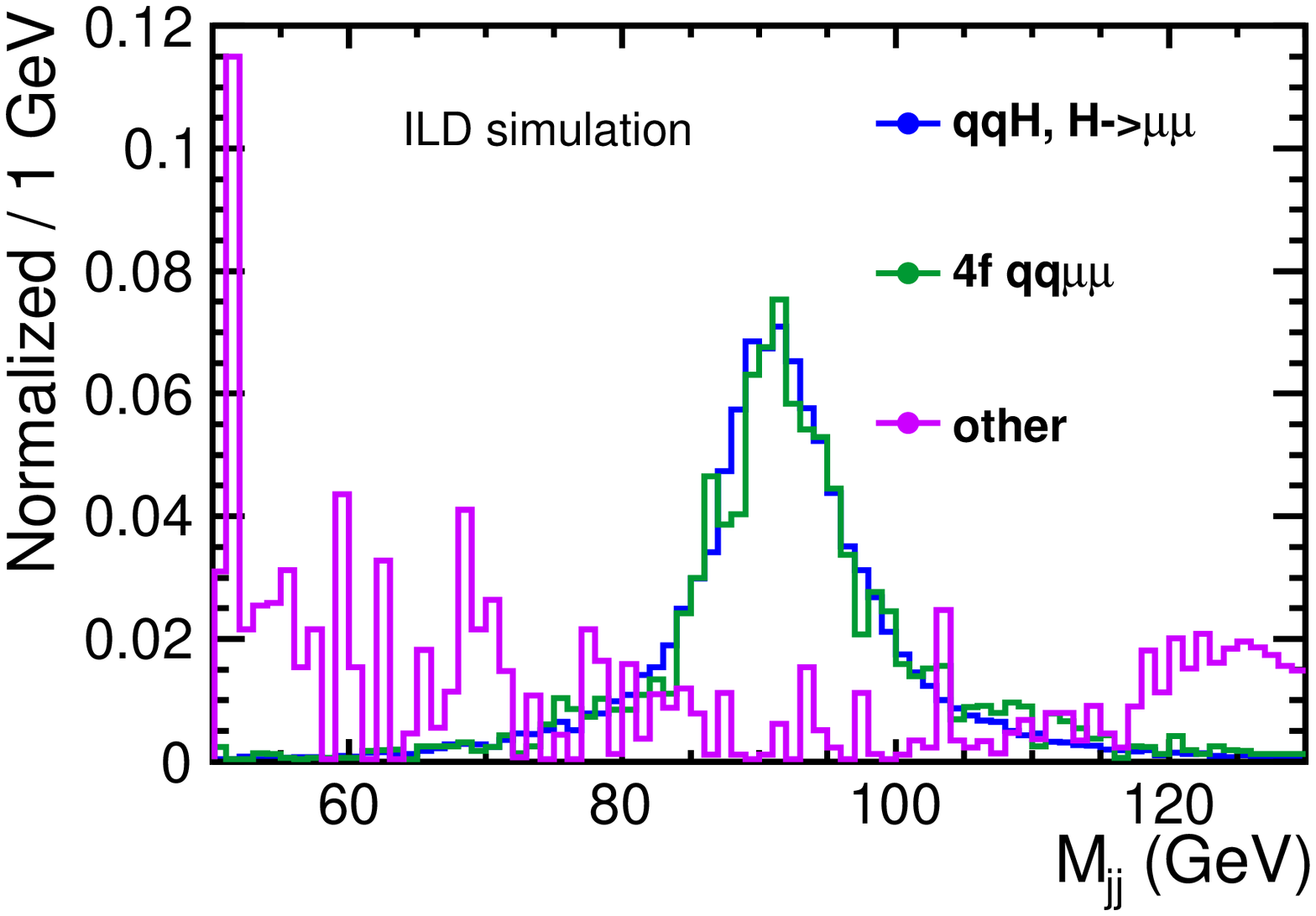}
  \caption{\label{fig:input1:mjj} $M_{jj}$}
\end{subfigure}
\begin{subfigure}{0.45\textwidth}
  \includegraphics[width = \textwidth]{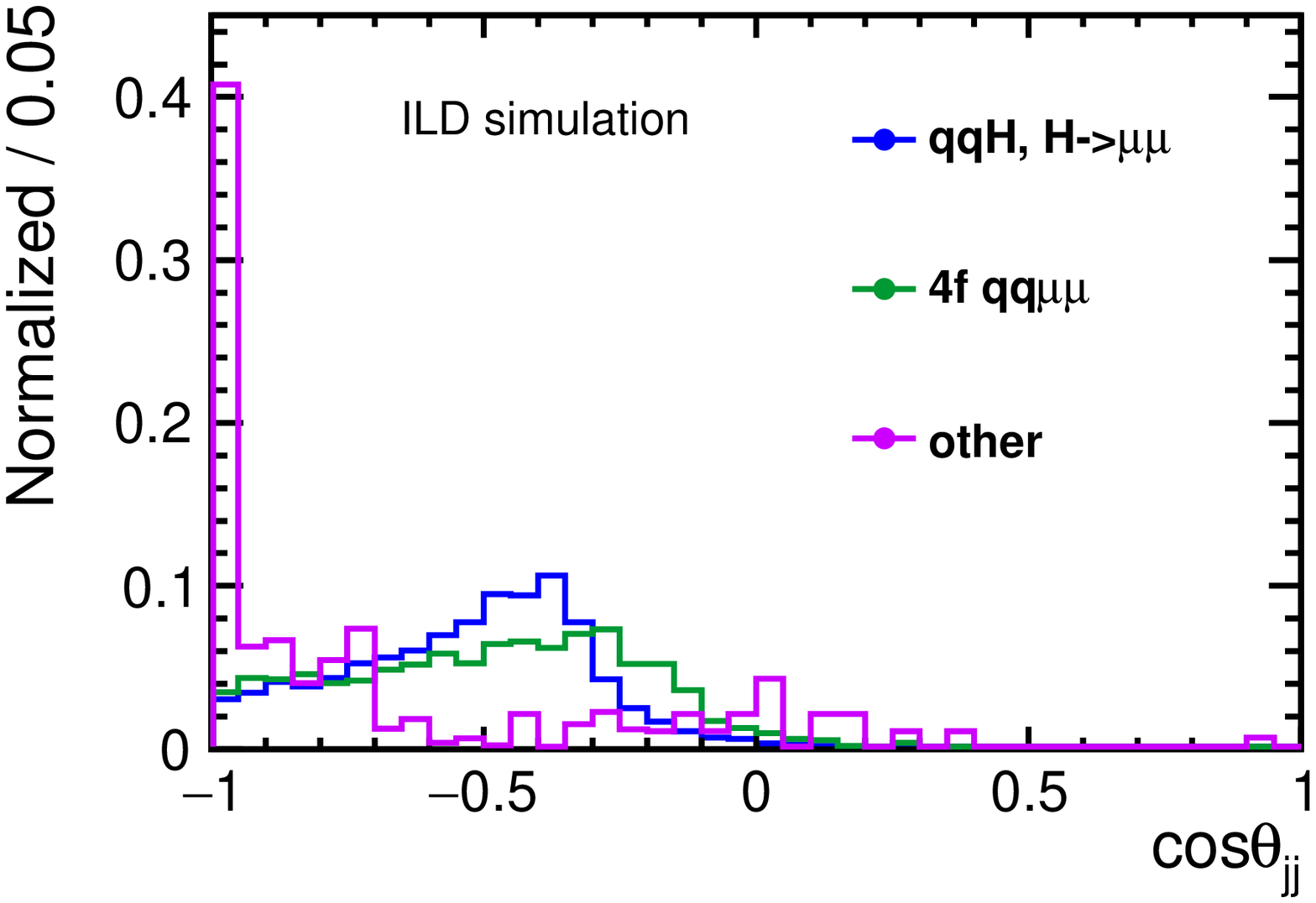}
  \caption{\label{fig:input1:ctjj} $\cos \theta _{jj}$}
\end{subfigure}
\begin{subfigure}{0.45\textwidth}
  \includegraphics[width = \textwidth]{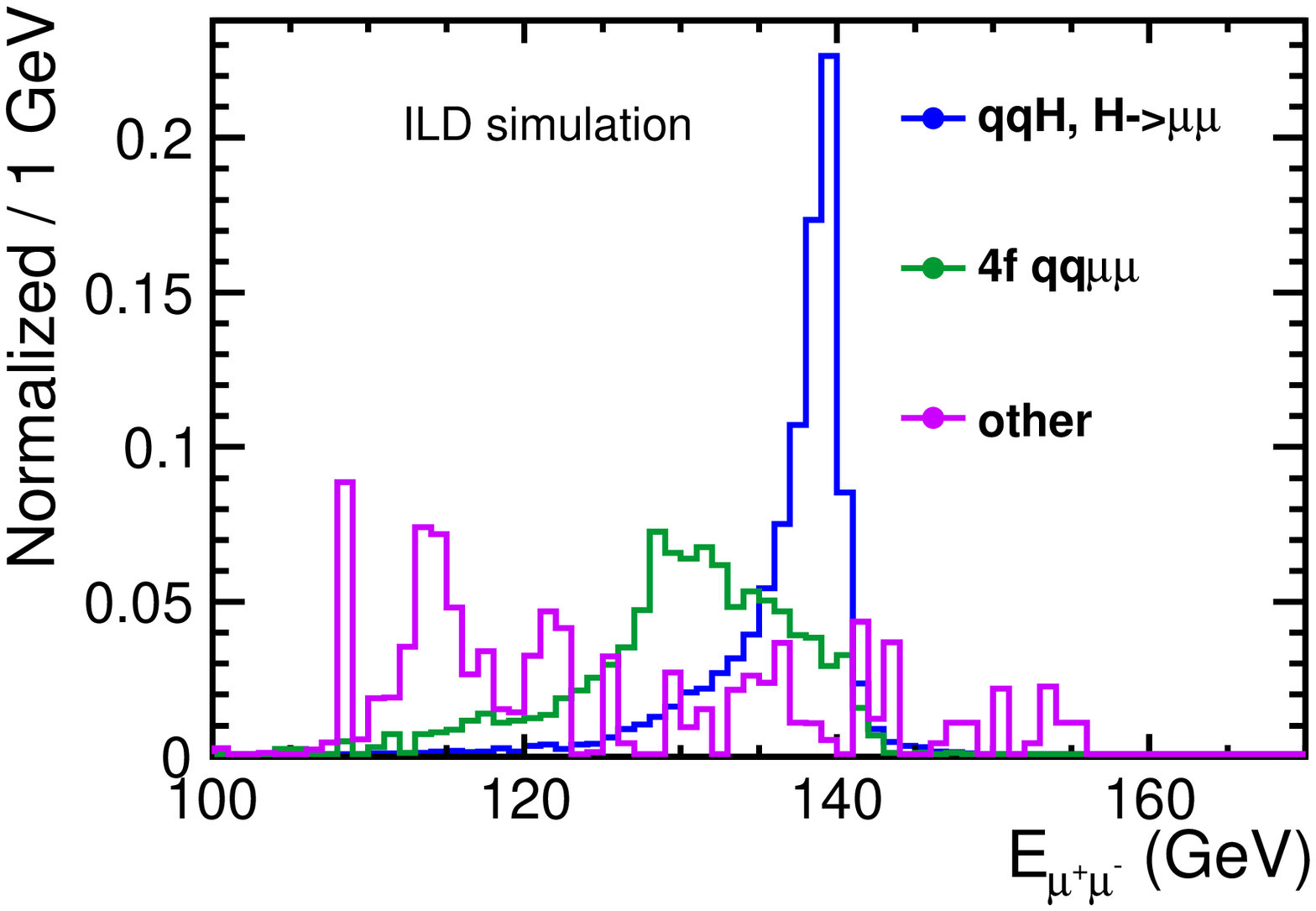}
  \caption{\label{fig:input1:Emm} $E_{\mu ^+ \mu ^-}$}
\end{subfigure}
\begin{subfigure}{0.45\textwidth}
 \includegraphics[width = \textwidth]{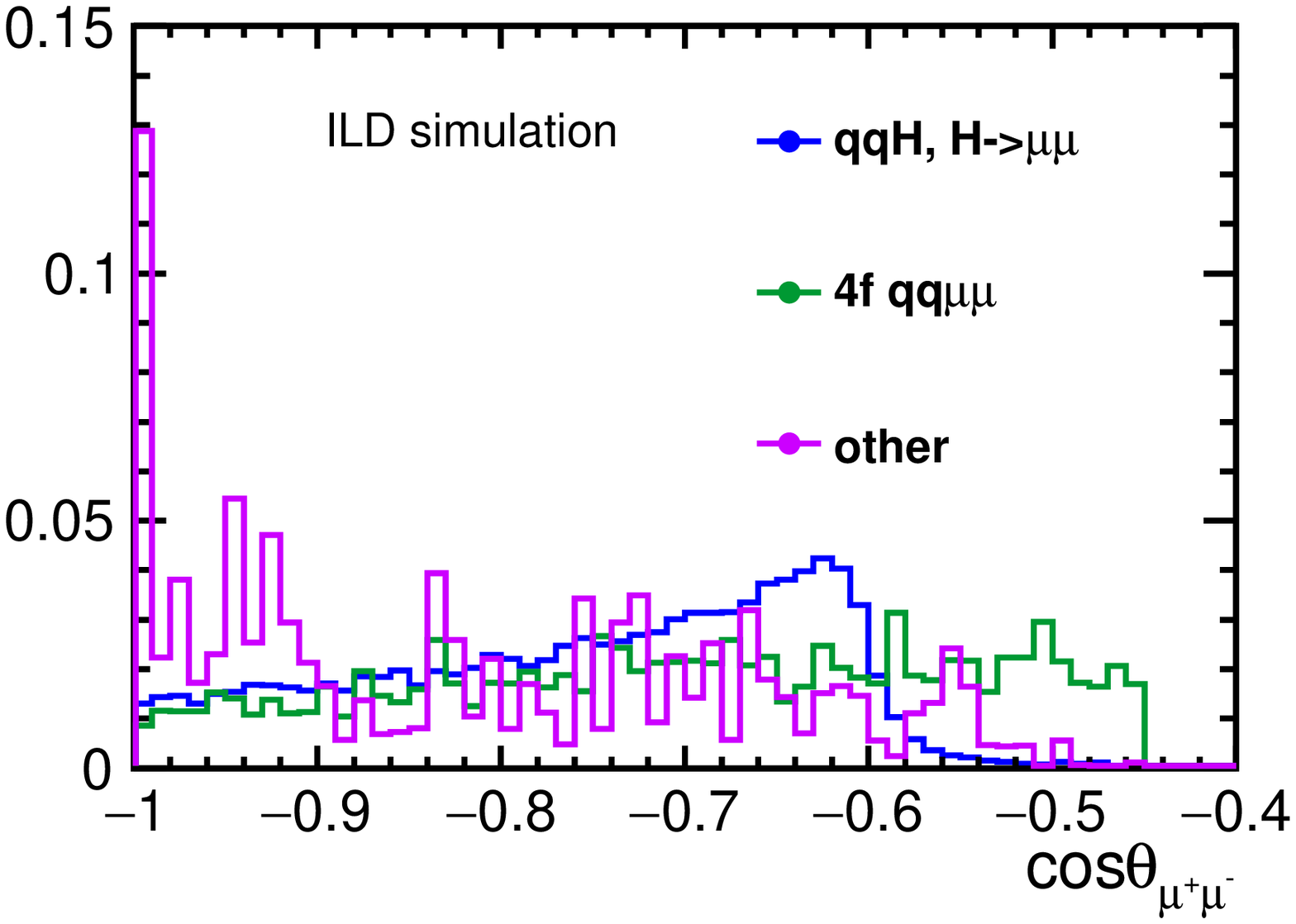}
 \caption{\label{fig:input1:ctmm} $\cos \theta _{\mu ^+ \mu ^-}$}
\end{subfigure}
\begin{subfigure}{0.45\textwidth}
  \includegraphics[width = \textwidth]{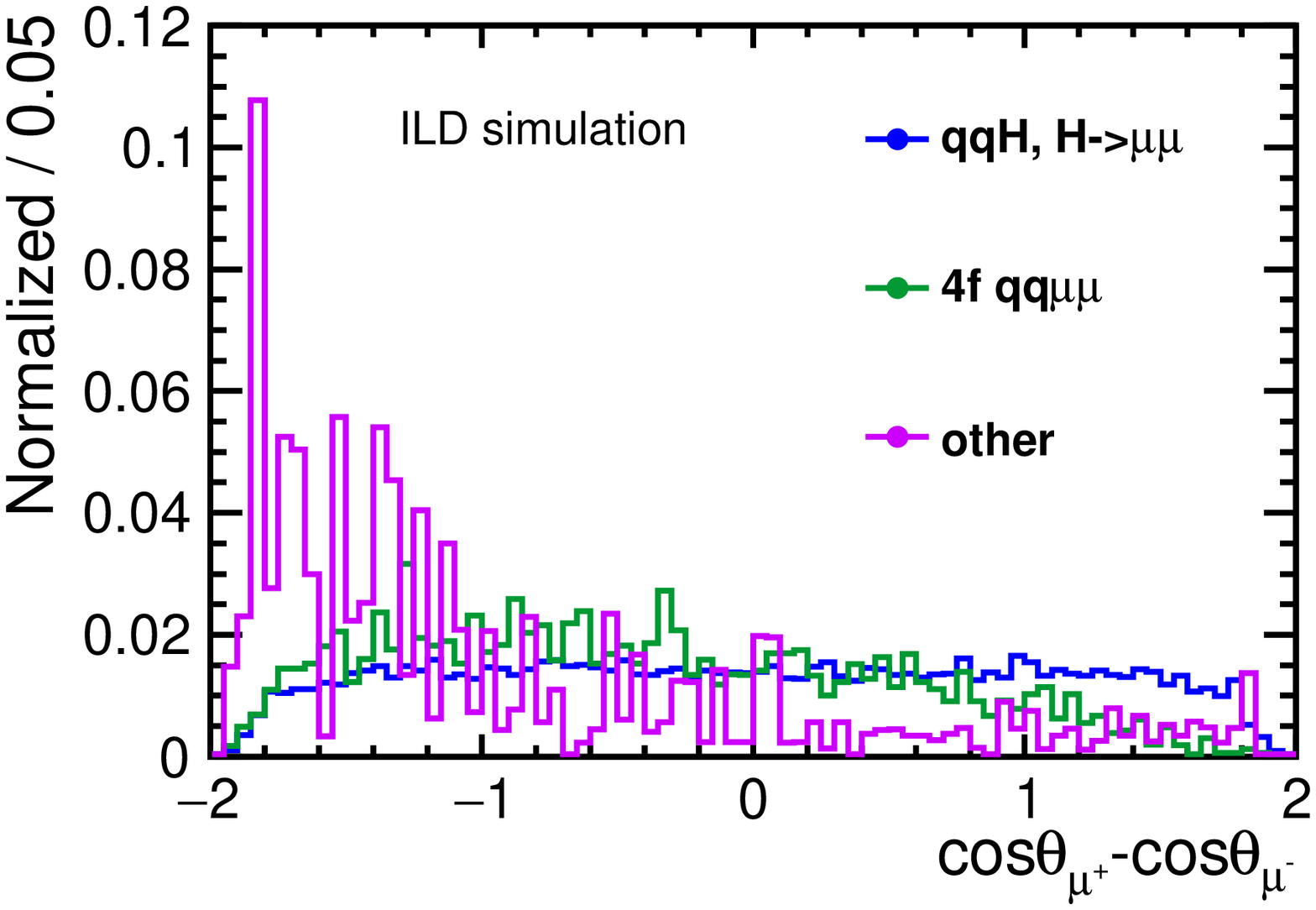}
  \caption{\label{fig:input1:scct} $\cos \theta _{\mu ^+} - \cos \theta _{\mu ^-}$}
\end{subfigure}
\begin{subfigure}{0.45\textwidth}
  \includegraphics[width = \textwidth]{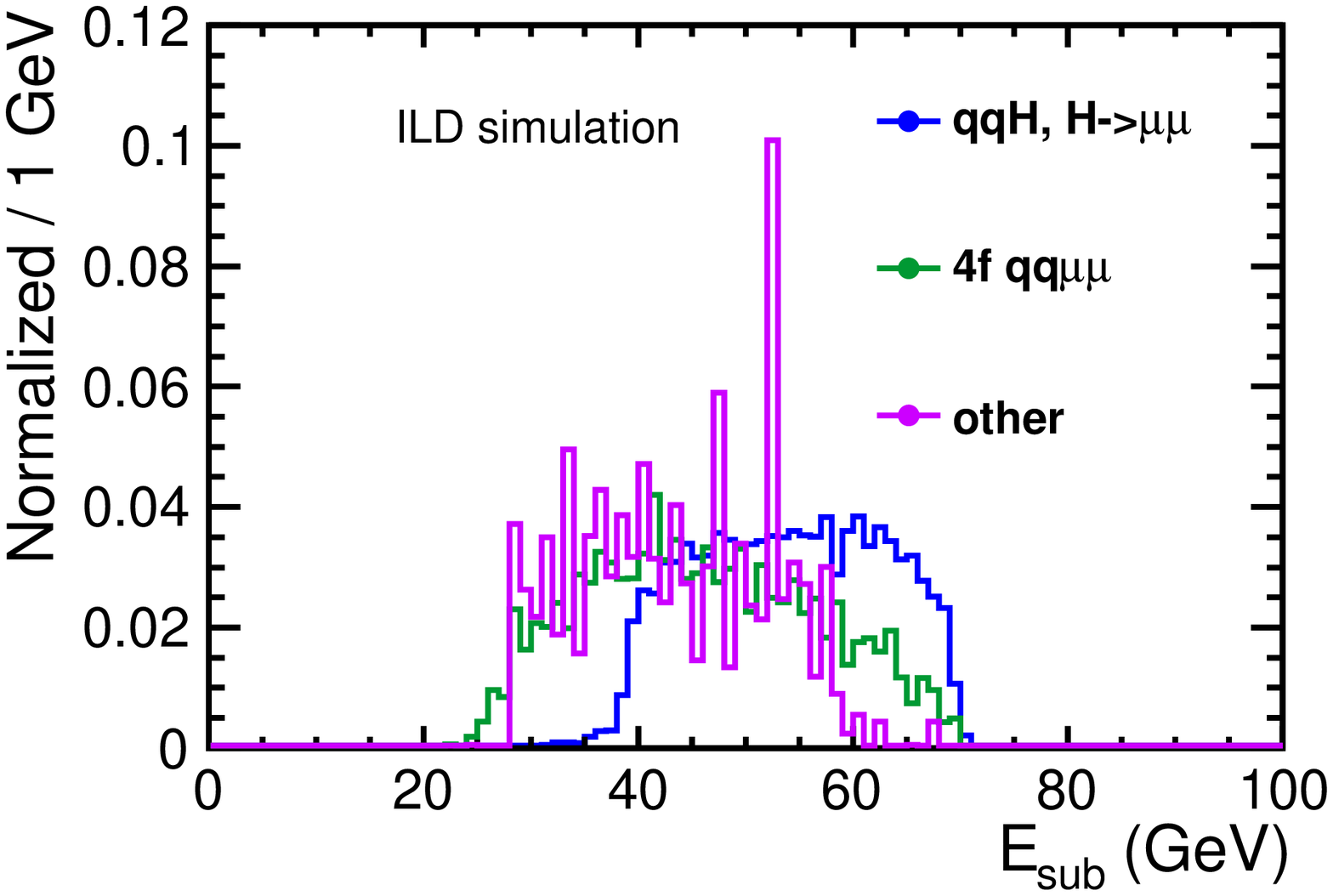}
  \caption{\label{fig:input1:Esl} $E_{\mathrm{sub}}$ }
\end{subfigure}
\begin{subfigure}{0.45\textwidth}
  \includegraphics[width = \textwidth]{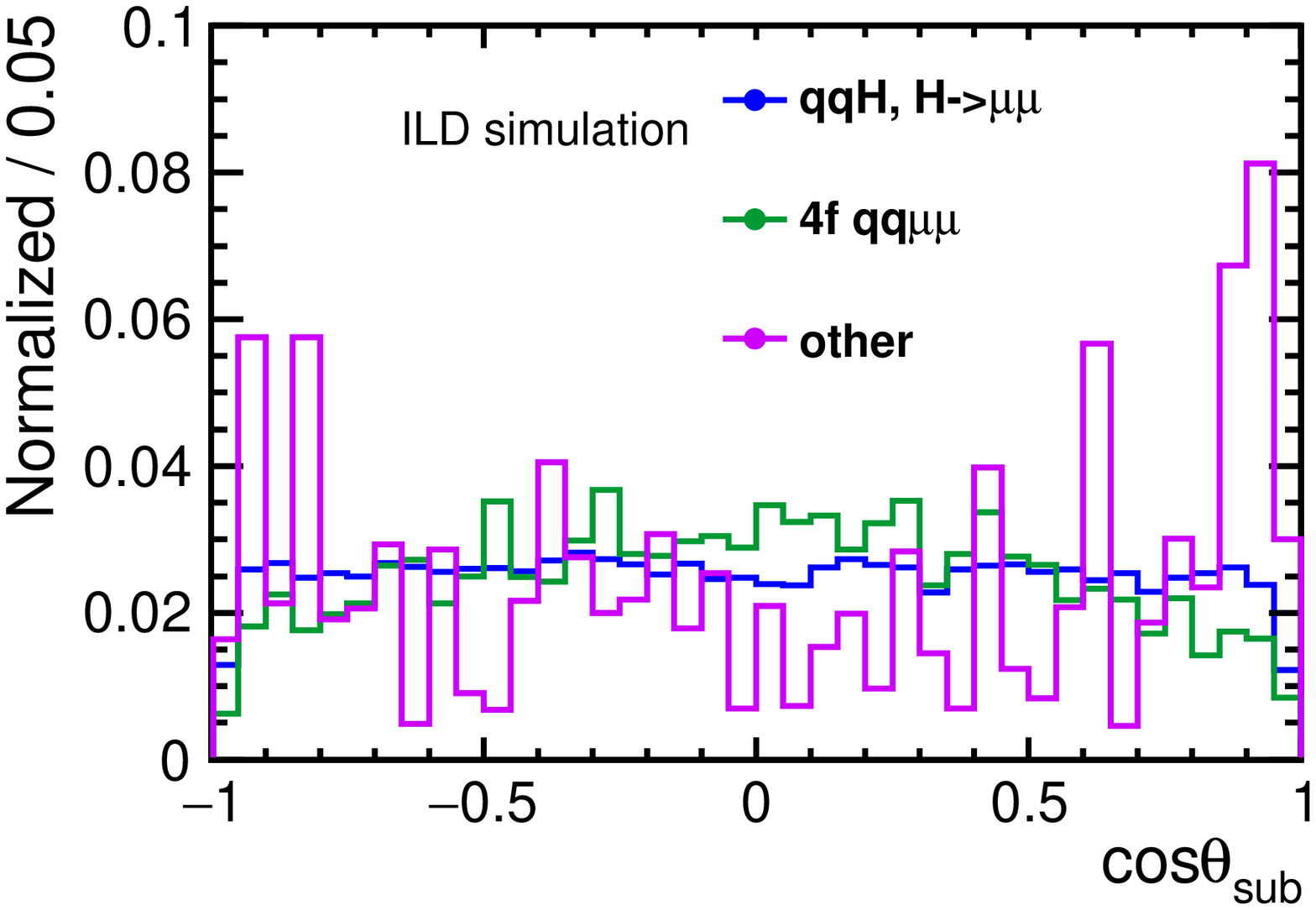}
  \caption{\label{fig:input1:ctsl} $\cos \theta _{\mathrm{sub}}$ }
\end{subfigure}
\end{center}
\caption{Example input variables for the BDTG in qqH250-L.
All histograms are normalised to an integral of 1.
The signal process is shown in blue, the irreducible backgrounds are contained in the green histograms, while other backgrounds are shown in magenta.}
\label{fig:input1}
\end{figure}

\begin{figure}[htb]
\centering
\includegraphics[width = 0.6\textwidth]{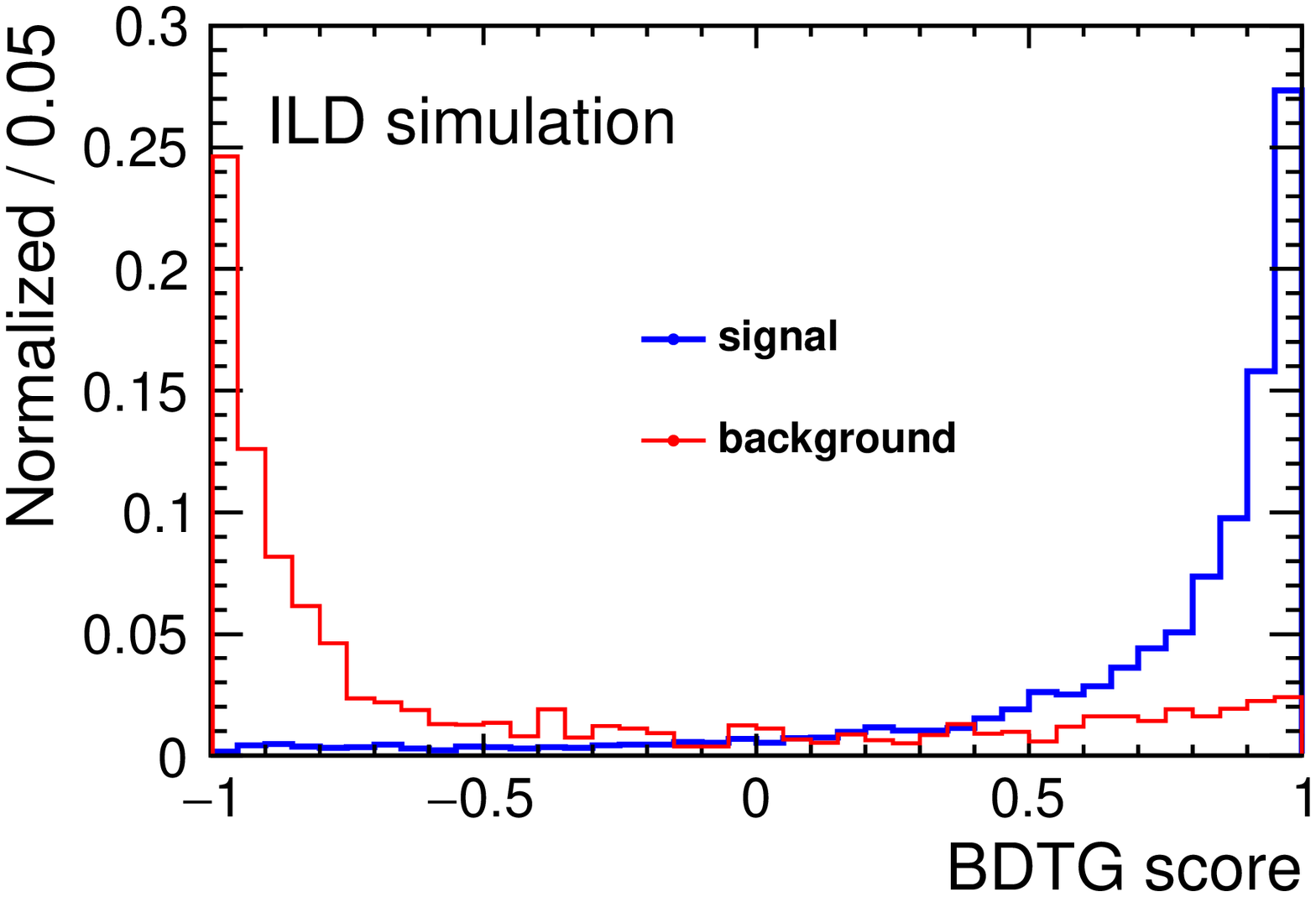}
\caption{The distribution of BDTG score (qqH250-L).
Both histograms are normalised to an integral of 1.}
\label{fig:BDTGscore}
\end{figure}

For the further rejection of background, in particular the ``irreducible'' one, a multivariate analysis is performed based on the gradient boosted decision tree method (BDTG) implemented in the TMVA package in ROOT~\cite{TMVA, ROOT}.
Typically, $\sim 10^4$ MC events remain after the preselection for signal and background, each. 
In all channels, half of the remaining events after the channel-specific preselection are used for training and the other half for testing.
The variable $M_{\mu ^+ \mu ^-}$ is not used in the BDTG since it will be used later in further analysis.
The input variables for the BDTG are summarised in Table~\ref{tab:inputvariables} for all the channels.
Here, $\theta _{jj}$ is the angle between two jets, $E_{\mu ^+ \mu ^-}$ is the energy sum of the $H \to \mu ^+ \mu ^-$ candidate, $\theta _{\mu ^+ (\mu ^-)}$ is the polar angle of $\mu ^+ (\mu ^-)$, $E_{\mathrm{lead}}(E_{\mathrm{sub}})$ is the energy of the higher energy (lower energy) muon of the $H \to \mu ^+ \mu ^-$ candidate, $\theta _{\mathrm{lead}}(\theta _{\mathrm{sub}})$ is the polar angle of the higher energy (lower energy) muon of the $H \to \mu ^+ \mu ^-$ candidate, $M_{\mathrm{recoil}}$ is the recoil mass against the $H \to \mu ^+ \mu ^-$ candidate (corrected for reconstructed ISR photons), $P_{t, \ \mu ^+ \mu ^-}$ is the transverse momentum of the $H \to \mu ^+ \mu ^-$ candidate system, and $\theta _ {\mathrm{thrustaxis}}$ is the polar angle of the thrust axis of the visible part of the event.
The variable $E_{\mu ^+ \mu ^-}$ is used in nnH500-L but not in nnH500-R.
For each channel, the minimum number of relevant inputs has been chosen, considering the minimisation of correlations and avoiding overtraining, in particular given the finite amount of available MC events.
Fig.~\ref{fig:input1} shows the seven input variables for qqH250-L, while the corresponding distribution of the BDTG score for the signal and background events is displayed in Fig.~\ref{fig:BDTGscore}.

\begin{table}[t]
\centering
\caption{Input variables to the BDTG for each channel.
The definition of variables is in the text.}
\begin{tabular}{c|l}
\hline
channel & input variables \\
\hline
qqH250-L/R & $M_{jj}$, $\cos \theta _{jj}$, \\
 & $E_{\mu ^+ \mu ^-}$, $\cos \theta _{\mu ^+ \mu ^-}$, $\cos \theta _{\mu ^+} - \cos \theta _{\mu ^-}$, \\
 & $E_{\mathrm{sub}}$, $\cos \theta _{\mathrm{sub}}$ \\
\hline
nnH250-L/R & $E_{\mathrm{vis}}$, $E_{\mu ^+ \mu ^-}$, $\cos \theta _{\mu ^+} - \cos \theta _{\mu ^-}$, $M_{\mathrm{recoil}}$, \\
 & $E_{\mathrm{sub}}$, $\cos \theta _{\mathrm{sub}}$ \\
\hline
qqH500-L/R & $M_{jj}$, $\cos \theta _{jj}$, \\
 & $P_{t, \mu ^+ \mu ^-}$, $\cos \theta _{\mu ^+ \mu ^-}$, $\cos \theta _{\mu ^+} - \cos \theta _{\mu ^-}$, $M_{\mathrm{recoil}}$, \\
 & $E_{\mathrm{lead}}$, $E_{\mathrm{sub}}$, $\cos \theta _{\mathrm{lead}}$, $\cos \theta _{\mathrm{sub}}$ \\
\hline
nnH500-L/R & $E_{\mathrm{vis}}$, $\cos \theta _{\mathrm{thrustaxis}}$, \\
 & $E_{\mu ^+ \mu ^-}$, $\cos \theta _{\mu ^+ \mu ^-}$, $\cos \theta _{\mu ^+} - \cos \theta _{\mu ^-}$, \\
 & $E_{\mathrm{lead}}$, $E_{\mathrm{sub}}$, $\cos \theta _{\mathrm{lead}}$, $\cos \theta _{\mathrm{sub}}$ \\
\hline
\end{tabular}
\label{tab:inputvariables}
\end{table}

For each channel, the final cut value on the BDTG score is chosen such that it optimises the expected precision on $\mathrm{BF}(H \to \mu^+\mu^-)$ as described in the following section. 


\subsection{Extraction of the Signal Strength}
\label{subsec:analysis:toymc}
After applying a cut on the BDTG score, the signal strength is extracted from a template fit to the invariant di-muon mass $M_{\mu ^+ \mu ^-}$ distribution for signal and background with the signal normalisation as a free parameter.
The final precision on the branching fraction is estimated via a toy MC technique.

First, the modeling functions for the $M_{\mu ^+ \mu ^-}$ distributions of signal and background have to be defined.
These functions are fitted to the $M_{\mu ^+ \mu ^-}$ distributions for signal and background as obtained from the full simulation analysis described in the previous subsections.

\begin{figure}[htbp]
\begin{center}
\begin{subfigure}{0.475\textwidth}
  \includegraphics[width = \textwidth]{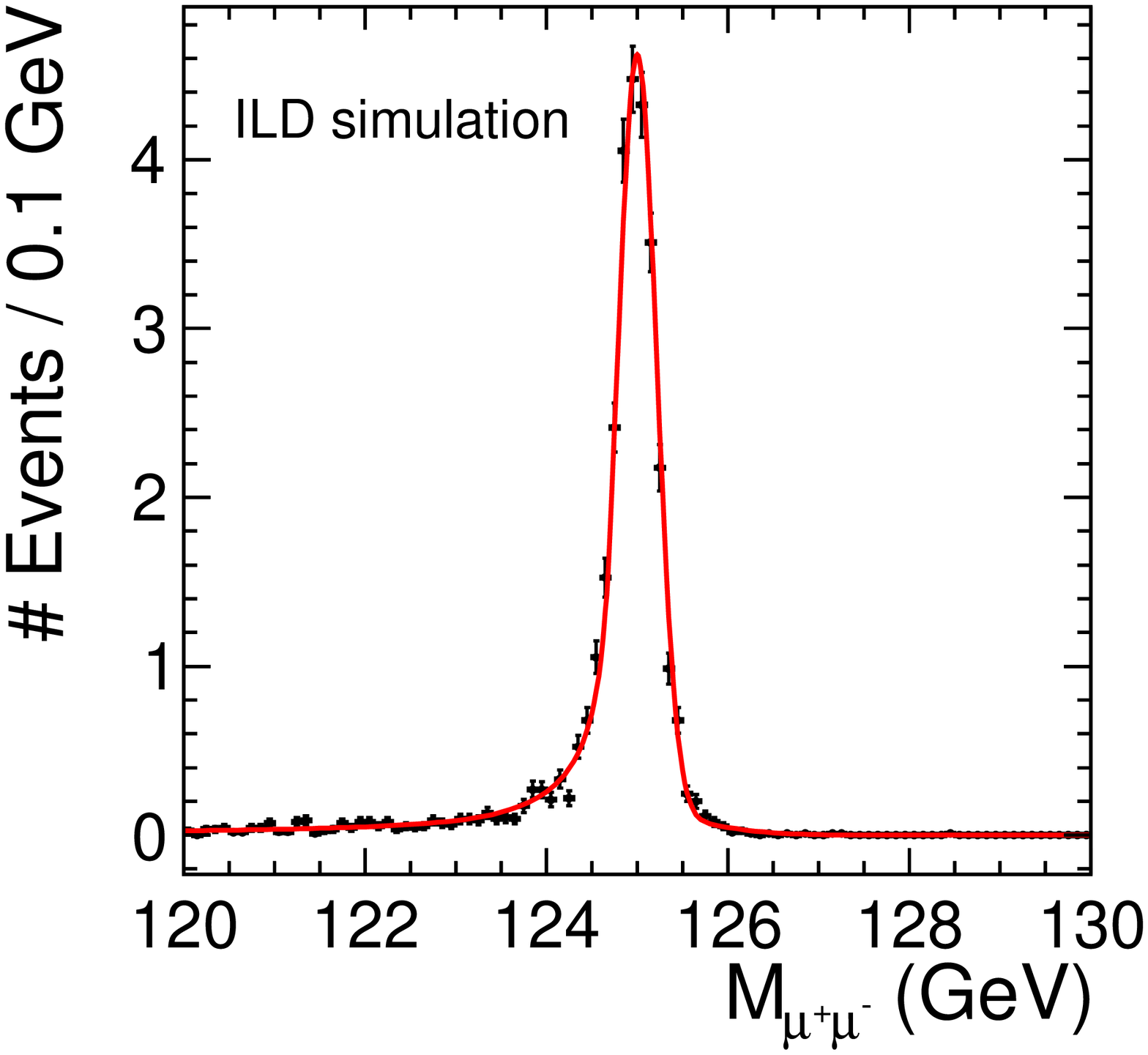}
  \caption{\label{fig:CBGfit_nnh500L:sig} signal with result of $f_S$ fit }
\end{subfigure}
\begin{subfigure}{0.475\textwidth}
  \includegraphics[width = \textwidth]{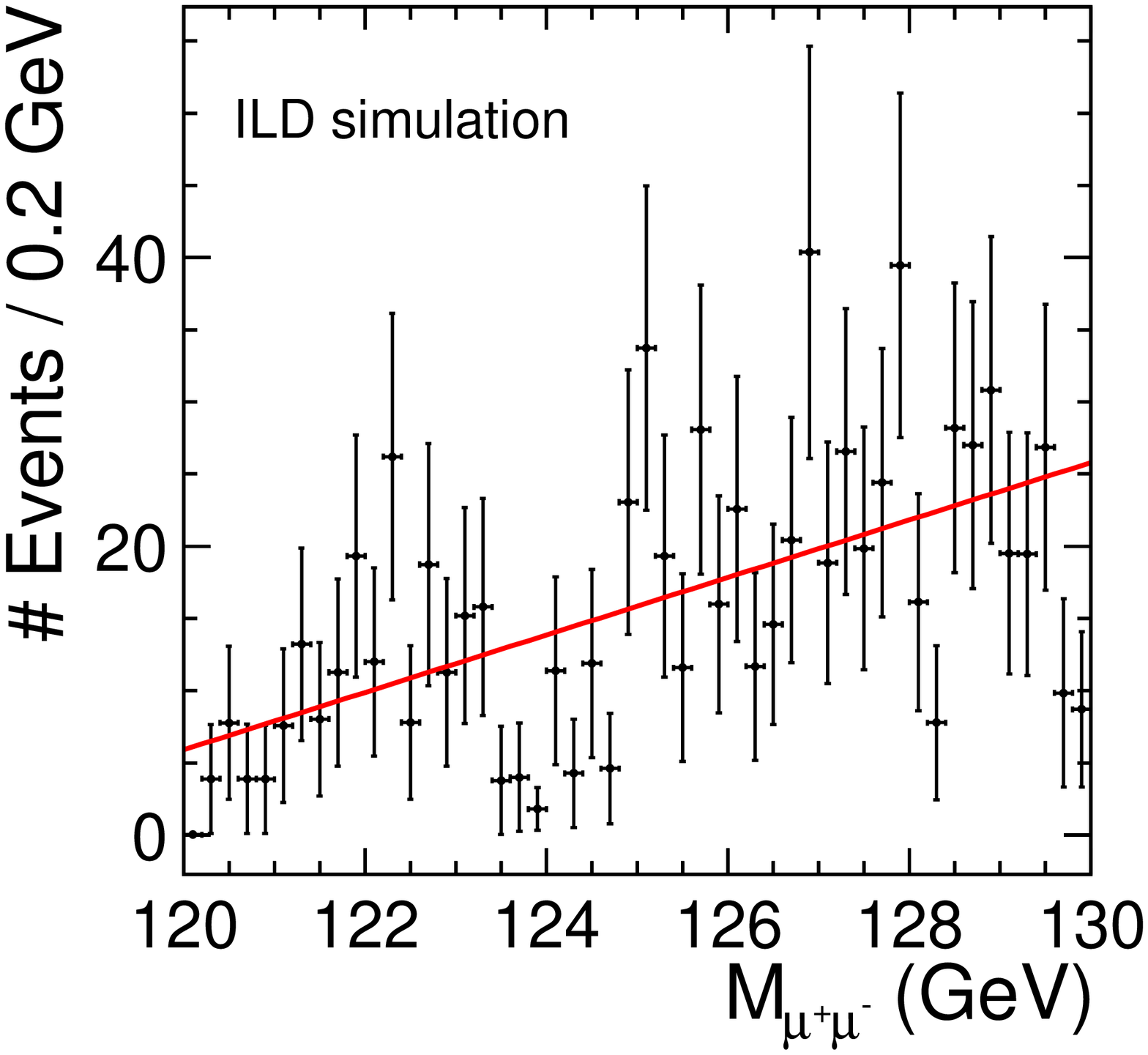}
  \caption{\label{fig:CBGfit_nnh500L:bkg} background with result of $f_B$ fit}
\end{subfigure}
\end{center}
\caption{$M_{\mu ^+ \mu ^-}$ distributions for signal and background after all cuts in the nnH500-L channel. The result of the $f_S$ and $f_B$ fits, respectively, is shown as red curves.}
\label{fig:CBGfit_nnh500L}
\end{figure}

Due to the excellent mass reconstruction, the whole template fit is restricted to the range $120\,\mathrm{GeV}<M_{\mu ^+ \mu ^-}< 130\,\mathrm{GeV}$. For the signal, a linear sum of a Crystal Ball function (CB)~\cite{CB} and a Gaussian,
\begin{equation}
f_S \equiv k \times \mathrm{CB} + (1-k) \times \mathrm{Gaussian} \quad (\mathrm{with\ } 0 < k < 1),    
\label{eq:model}
\end{equation}
is used as modeling function $f_S$.
This empirical function models sufficiently well the combined effect of final state radiation photons, which create a tail in the $M_{\mu ^+ \mu ^-}$ distribution of the signal process, as well as effects of the finite detector resolution.
It should be noted that no attempt has been made to recover final state radiation photons because the measuring accuracy of the electromagnetic calorimeter is not good enough to improve the mass resolution of events with recovered photons.
As we will see in Sec.~\ref{sec:impact:ptres}, an excellent invariant mass resolution is a core ingredient to the final performance of the analysis, and thus recovery of final state radiation is not considered worthwhile in this case.
For the signal modeling, an unbinned fit is performed to avoid effects of the bin width which, due to finite MC statistics, cannot always be small compared to the width of the mass peak, especially when considering different $P_t$ resolutions in Sec.~\ref{sec:impact:ptres}.
The Higgs mass itself is assumed to be known very precisely, to about $14$\,MeV, from the recoil analysis~\cite{LeptonicRecoil}.
Therefore, the mean value of the CB is fixed to the nominal Higgs mass of 125\,GeV in this study.
Figure~\ref{fig:CBGfit_nnh500L:sig} illustrates the modeling of the signal $M_{\mu ^+ \mu ^-}$ distribution using the nnH500-L channel as example. 
In this example, the parameter $k$ in Eq.~(\ref{eq:model}) is 0.92. The width of the peak at half its maximum height (FWHM) is 0.23 GeV.

The background is modeled by a straight line $f_B$ in all channels.
An example is given in Fig.~\ref{fig:CBGfit_nnh500L:bkg}, again based on the nnH500-L channel. 


\begin{figure}[htbp]
\begin{center}
\begin{subfigure}{0.45\textwidth}
  \includegraphics[width = \textwidth]{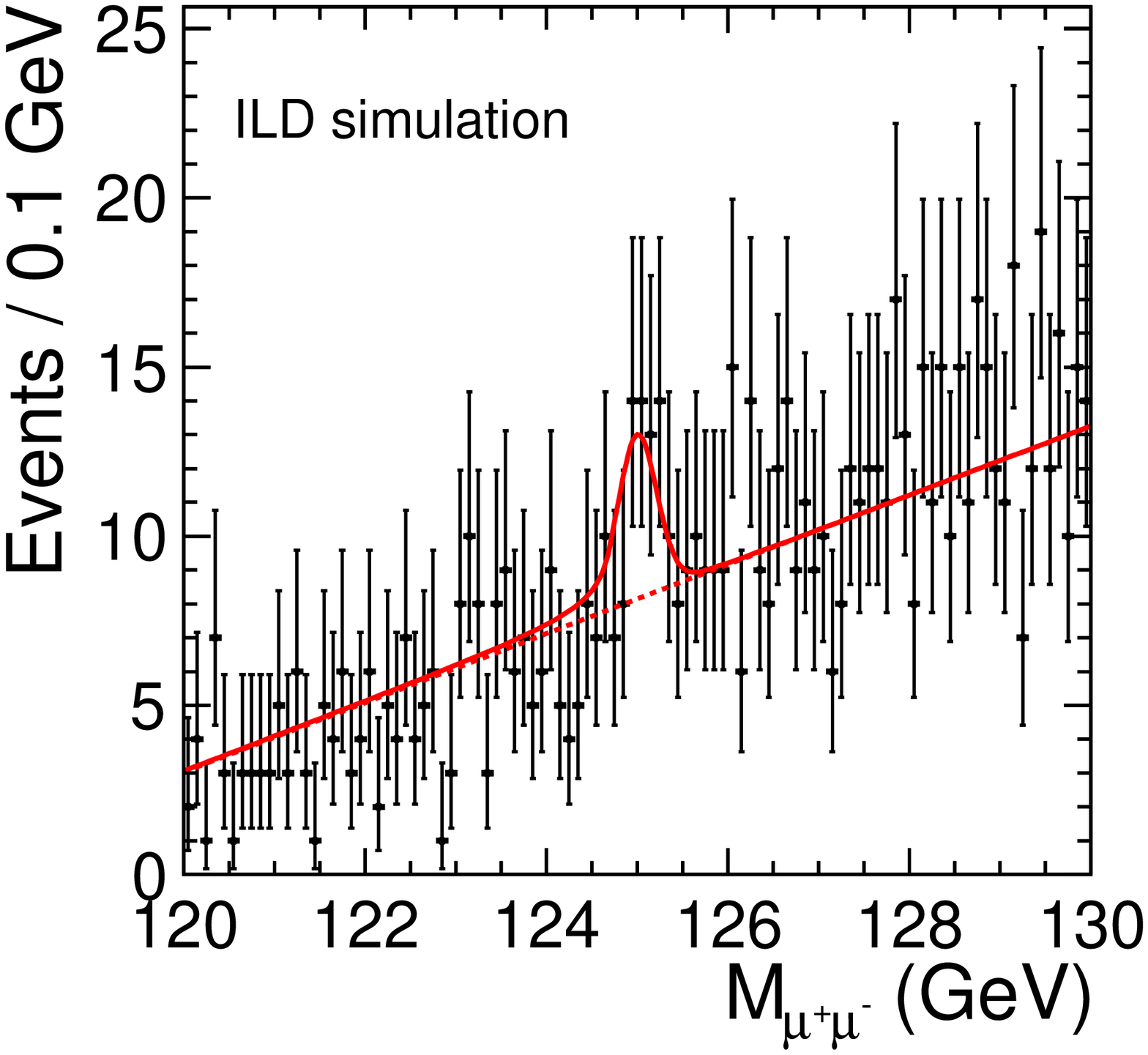}
  \caption{\label{fig:toyMC_nnh500L:1} example of one pseudo-experiment}
\end{subfigure}
\begin{subfigure}{0.45\textwidth}
  \includegraphics[width = \textwidth]{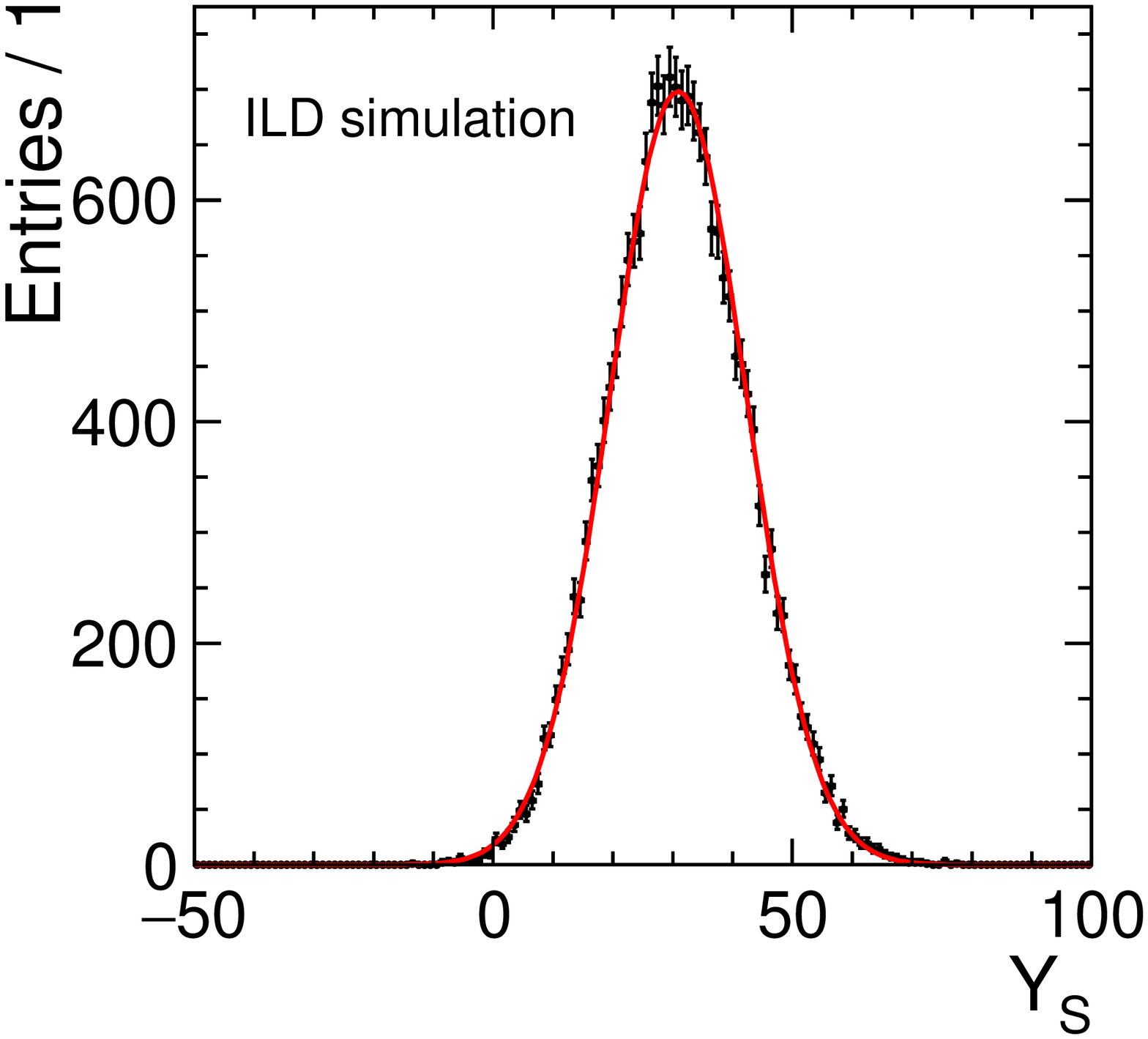}
  \caption{\label{fig:toyMC_nnh500L:2} distribution of yield of signal events $Y_S$ from $2 \times 10^4$ pseudo-experiments.}
\end{subfigure}
\end{center}
\caption{Signal strength extraction in the nnH500-L channel.
(a) Example outcome of one pseudo-experiment, with the pseudo-data shown as black points with error bars, while the solid red curve shows the result of an unbinned fit of $f \equiv Y_Sf_S + Y_Bf_B$ to the pseudo-data.
The dashed red line shows its background component $Y_Bf_B$.
(b) The distribution of the yield of signal events $Y_S$ obtained from $2 \times 10^4$ pseudo-experiments, fitted with a Gaussian function.
The mean value of the distribution is $Y_S=30.9$ with a width of 11.4 events, both obtained from the fitted Gaussian.}
\label{fig:toyMC_nnh500L}
\end{figure}

The fitted $f_S$ and $f_B$ are then used as probability density functions for the generation of $2\times 10^4$ pseudo-data sets via a toy MC technique based on RooFit~\cite{RooFit, ROOT}.
In each pseudo-experiment, the number of pseudo-signal(-background) events is drawn from a Poisson distribution with the estimated average number of signal(background) events after all cuts as expectation value. Then, an unbinned fit of the function $f \equiv Y_Sf_S + Y_Bf_B$ to the sum of pseudo-data is performed, where $Y_S(Y_B)$ is the yield of signal(background) events.
Thereby, $Y_B$ is fixed to the expected average number of backgrounds after all cuts, assuming that by the time the ILC has collected its full data set, the SM background at a lepton collider can be predicted much more precisely than the statistical uncertainty for rare signal events.
Thus, $Y_S$ is the only free parameter in the template fit.
Fig.~\ref{fig:toyMC_nnh500L} shows an example of one pseudo-experiment in the nnH500-L channel. The final $Y_S$ distribution from $2\times 10^4$ pseudo-experiments is fitted by a Gaussian to extract its mean and width as shown in Fig.~\ref{fig:toyMC_nnh500L}(b).
The expected relative precision on $\mathrm{BF}(H \to\mu ^+ \mu ^-)$ is calculated as the width of the fitted Gaussian divided by the mean of the fitted Gaussian, which in all channels agrees with the mean number of signal events expected from the full simulation listed in Table~\ref{tab:BDTGanalysis}. 


The cut values for the BDTG score have been optimised for each analysis channel by applying the toy MC procedure described above for different values of the cut and selecting the one which gives the best measurement precision.
Figs.~\ref{fig:CBGfit_nnh500L} and~\ref{fig:toyMC_nnh500L} correspond to the optimal BDTG score cut case in nnH500-L.

\subsection{Results and Discussion}
\label{subsec:analysis:result}
Table~\ref{tab:BDTGanalysis} shows the number of signal and background events in each channel after the optimisation of the BDTG score cut.
The signal efficiency ranges between $45\%$ and $72\%$, with an overall average of $53\%$.
A notable exception is the nnH250-L channel, which gives an optimal result for a very hard cut on the BDTG score and as a result has a rather low efficiency of only $28\%$, while the background is still higher than in its sister channel nnH250-R.
This is an effect of the $W^+W^-$ contribution to the irreducible background, which has a much higher cross-section in the left-handed polarisation configuration, while the corresponding increase for the signal is much smaller since it is --- at this energy --- dominated by $ZH$ production.
At $500$\,GeV, the effect on the background is even more drastic, but since the now $WW$-fusion dominated signal profits in the same way from the polarisation, there is no need to optimise for an extremely hard cut on the BDTG score. 
In all cases, the total background count is strongly dominated by the irreducible component.
This implies that the misidentification of the final-state particles is not a limiting factor in the analysis.
In the future it could be investigated, however, whether the event kinematics could be exploited in a more efficient way to suppress the irreducible component, as will be discussed in more detail below. 

\begin{table}[t]
\centering
\caption{The number of signal and background events in each channel after all cuts.
The optimal cut on the BDTG score is listed in the second column.
The numbers in brackets show the signal selection efficiency.}
\begin{tabular}{c|c|cccc}
\hline
\multirow{2}{*}{channel} & BDTG & \multirow{2}{*}{signal} & other & \multirow{2}{*}{``irreducible''} & other SM \\
 & score cut & & Higgs & & background \\
\hline
qqH250-L & $> 0.45$ & 29 (72{\%}) & 0.1 & 600 & 4  \\
qqH250-R & $> 0.85$ & 18 (64{\%}) & 0 & 193 & 3 \\
nnH250-L & $> 0.95$ & 4.2 (28{\%}) & 0 & 155 & 12 \\
nnH250-R & $> 0.80$ & 3.7 (45{\%}) & 0 & 105 & 11 \\
qqH500-L & $> 0.60$ & 13 (54{\%}) & 4.2 & 114 & 9 \\
qqH500-R & $> 0.25$ & 10 (61{\%}) & 9.6 & 71 & 7 \\
nnH500-L & $> 0.50$ & 31 (54{\%}) & 0 & 745 & 48 \\
nnH500-R & $> 0.40$ & 3.6 (45{\%}) & 0 & 75 & 1 \\
\hline
\end{tabular}
\label{tab:BDTGanalysis}
\end{table}


The expected precisions on $\mathrm{BF}(H \to\mu ^+ \mu ^-)$ obtained in the eight channels are summarised in Table~\ref{tab:results}.
With the $\sqrt{s}=250$\,GeV data alone, a precision of $23\%$ can be reached, dominated by the $q\bar{q}H$ channels.
The $\sqrt{s}=500$\,GeV data alone reach $24\%$, but now the $\nu \overline{\nu}H$ channel in the left-handed data set is the most sensitive.
A combination of all data sets improves the expected precision to $17\%$.

These numbers demonstrate a significant improvement with respect to earlier analyses.
An extrapolation of the result reported by SiD~\cite{mu5} to a luminosity of 0.9\,ab$^{-1}$, which corresponds to the size of the left-handed data set at $\sqrt{s} =250$\,GeV in the present study, yields a precision on $\mathrm{BF}(H \to\mu ^+ \mu ^-)$ of $\sim 48{\%}$.
This can be compared to the qqH250-L result of 34{\%} in the present analysis. This difference originates partially from a more sophisticated, multivariate analysis instead of the cut-based selection in the SiD study.
But more importantly, for the intermediate momentum range of $40$ to $100$\,GeV which is of relevance here, the ILD momentum resolution is considerably better than that of SiD: in the case of SiD~\cite{TDR4}, $\sigma _{1/P_t}$ ranges between $3 \times 10^{-5}$\,GeV$^{-1}$ for $P=100$\,GeV at $\theta=90^{\circ}$ to $2 \times 10^{-4}$\,GeV$^{-1}$ for $P=40$\,GeV at $\theta=30^{\circ}$, while in the ILD case the relevant numbers range from $2 \times 10^{-5}$\,GeV$^{-1}$ for $P=100$\,GeV at $\theta=85^{\circ}$ to $5 \times 10^{-5}$\,GeV$^{-1}$ for $P=40$\,GeV at $\theta=30^{\circ}$, as shown in Fig.~\ref{fig:DBD_momres}. 

\begin{table}[t]
\centering
\caption{Expected precisions on $\mathrm{BF}(H \to \mu ^+ \mu ^-)$ for $\sqrt{s}=250$\,GeV (ILC250), $\sqrt{s}=500$\,GeV (ILC500) and their combination (ILC250+500).
The luminosities and polarisation sharing correspond to the standard ILC running scenario as detailed in Sec.~\ref{sec:intro}.}
\begin{tabular}{ccc|c|c} \hline
$\sqrt{s} = 250$ GeV & $q\overline{q}H$ & $\nu \overline{\nu} H$ & ILC250 & ILC250+500\\
\hline
L & 34{\%} & 113{\%} & \multirow{2}{*}{23{\%}} & \multirow{5}{*}{17{\%}}\\
R & 36{\%} & 111{\%} & \\
\cline{1-4}
$\sqrt{s} = 500$ GeV & $q\overline{q}H$ & $\nu \overline{\nu} H$ & ILC500 \\
\cline{1-4}
L & 43{\%} & 37{\%} & \multirow{2}{*}{24{\%}} \\
R & 48{\%} & 106{\%} & \\
\hline
\end{tabular}
\label{tab:results}
\end{table}

The combined result of the present study is about $50\%$ 
larger than the most recent projections for the HL-LHC based on the tracker upgrades for ATLAS and CMS introduced in Sec.~\ref{sec:intro}.
Taking into account that HL-LHC will provide $O(10^4)$ $H \to \mu ^+ \mu ^-$ events with the full expected integrated luminosity of 3\,ab$^{-1}$, and thus about $100$ times more signal events than ILC with 2\,ab$^{-1}$ at $\sqrt{s} =250$\,GeV and 4\,ab$^{-1}$ at $\sqrt{s} =500$\,GeV together, a difference of only $50\%$ shows the highly efficient use of data possible at an $e^+e^-$ collider.
In addition, the results of the analysis at ILC1000 presented in Ref.~\cite{mu2} can be extrapolated to the full luminosity of 8\,ab$^{-1}$ expected at ILC1000~\cite{ILCOperatingScenario} to yield a precision of 14{\%} on $\mathrm{BF}(H \to \mu ^+ \mu ^-)$.
Combining this result with our analysis at 250\,GeV and 500\,GeV yields a precision of 11{\%}.
Thus, from a combination of HL-LHC with the full ILC program a precision of 7{\%} could be expected, without taking into consideration possible improvements of the analysis.

For a better understanding of analysis limitations, one can compare these results with the ``theoretical limit'' case, \textit{i.e.} assuming 100{\%} signal selection efficiency and no backgrounds.
In this hypothetical case, the precision would reach 10.4{\%} for ILC250, and 7.1{\%} for the full ILC250+500 data set.
The results presently achieved in full detector simulation are about a factor of $2.4$ more than these theoretical limits for three reasons: the signal efficiency of $53\%$ on average, the remaining irreducible backgrounds, and the invariant mass resolution for the di-muon system. 
\begin{itemize}
    \item If only the signal efficiencies as given in Table~\ref{tab:BDTGanalysis} are considered, the combined precision at ILC250 would be 13.4{\%} and would improve to 9.4{\%} when combined with the $500$~GeV data, which is a factor of $\sim 1.7$ better than the full simulation results.
    Table~\ref{tab:DetailsCutFlow} shows the detailed cut flow of the single most sensitive channel qqH250-L as an example.
    At the ``common cuts'' stage, $\sim 10\%$ of the signal events are lost.
    In about $4\%$ of the events, the muons are not found, and about $2.5\%$ of events are lost due to the $d_0$ and invariant mass requirements, each, c.f.; Table~\ref{tab:DetailsCutFlow}.
    The invariant mass requirement mostly fails due to the presence of FSR, which is not recovered, c.f.\ discussion in Sec.~\ref{subsec:analysis:toymc}.
    During the rest of the preselection, only a few additional percents are lost, while a $\sim 15\%$ reduction occurs via the BDTG score cut.
    In total, it seems hard to increase the overall signal efficiency drastically, but some improvement could be achieved by exploiting the variables discussed in the next item.
   \item The irreducible background almost entirely consists of processes with the same final state as the signal process: $e^+ e^- \to qq\mu ^+ \mu ^-$ for the $q\overline{q}H$ process and $e^+ e^- \to \nu \nu \mu ^+ \mu ^-$ for the $\nu \overline{\nu} H$ process, originating from $ZZ$ as well as, in the case of $\nu \nu \mu ^+ \mu ^-$, from $W^+W^-$ production and single-$Z/\gamma$ radiation off a $t$-channel $W$. 
   Future upgrades of the analyses could attack these kinds of backgrounds by even better exploitation of all kinematic information, \textit{e.g.} by testing various intermediate boson hypotheses in a kinematic fit~\cite{Beckmann:2010ib} and/or by evaluating the matrix element probabilities for the signal and various background hypotheses on an event-by-event basis~\cite{Abbiendi:2000ug}.
   There are also some background events with tau leptons such as $\nu \nu \tau \mu$ with $\tau$ decaying to $\mu$, but this contribution is negligible compared to the ones with exactly the signal final state.
   \item Last but not least, the di-muon invariant mass resolution is an important ingredient -- as soon as the number of background events at the end of the selection is larger than zero: the sharper the reconstructed Higgs mass peak, the lower the ``effective'' amount of background under the peak.
   The invariant mass resolution is dominated by the precision achieved on the transverse momentum of the two muons, while the angular resolutions play a negligible role.
   Therefore, we will study the impact of the (inverse) transverse momentum resolution $\sigma _{1/P_t}$ in Sec.~\ref{sec:impact:ptres}.
\end{itemize}

\begin{table}[t]
\centering
\caption{Detailed cut flow of the qqH250-L channel.
The numbers in brackets show the signal selection efficiency.}
\begin{tabular}{cc|cccc}
\hline
\multirow{2}{*}{} & & \multirow{2}{*}{signal} & other & \multirow{2}{*}{``irreducible''} & other SM \\
 & & & Higgs & & background \\
\hline
\multicolumn{2}{c|}{no cut} & 41 (100{\%}) & $2.9 \times 10^{5}$ & $2.96 \times 10^{5}$ & $1.8 \times 10^{9}$  \\ \hline
\multicolumn{2}{c|}{{\#} $\mu ^{\pm} = 1$} & 40 (96{\%}) & $9.5 \times 10^{3}$ & $1.12 \times 10^{5}$ & $7.3 \times 10^{7}$ \\ \hline
\multirow{6}{*}{\shortstack{common cuts \\ (see Table~\ref{tab:hmumucuts})}} & {\#}1 & 39 (95{\%}) & $9.4 \times 10^{3}$ & $1.10 \times 10^{5}$ & $6.5 \times 10^{7}$ \\
& {\#}2 & 38 (93{\%}) & $9.0 \times 10^{3}$ & $1.04 \times 10^{5}$ & $4.5 \times 10^{7}$ \\
& {\#}3 & 38 (93{\%}) & $9.0 \times 10^{3}$ & $1.04 \times 10^{5}$ & $4.5 \times 10^{7}$ \\
& {\#}4 & 38 (93{\%}) & $9.0 \times 10^{3}$ & $1.03 \times 10^{5}$ & $4.5 \times 10^{6}$ \\
& {\#}5 & 37 (90{\%}) & $2.0 \times 10^{2}$ & $5.08 \times 10^{3}$ & $3.1 \times 10^{5}$ \\
& {\#}6 & 37 (90{\%}) & $2.0 \times 10^{2}$ & $3.80 \times 10^{3}$ & $1.9 \times 10^{5}$ \\ \hline
\multirow{4}{*}{\shortstack{preselection \\ (see Table~\ref{tab:qqhcuts})}} & {\#}1 & 37 (90{\%}) & $1.9 \times 10^{2}$ & $3.79 \times 10^{3}$ & $1.7 \times 10^{5}$ \\
& {\#}2 & 37 (90{\%}) & $1.8 \times 10^{2}$ & $3.79 \times 10^{3}$ & $3.9 \times 10^{4}$ \\
& {\#}3 & 36 (89{\%}) & $1.7 \times 10^{2}$ & $3.71 \times 10^{3}$ & $2.2 \times 10^{3}$ \\
& {\#}4 & 36 (88{\%}) & $1.4 \times 10^{2}$ & $3.54 \times 10^{3}$ & $3.4 \times 10^{2}$ \\ \hline
\multicolumn{2}{c|}{BDTG score} & 30 (73{\%}) & 0.2 & 687 & 8.5  \\ \hline
\multicolumn{2}{c|}{$M_{\mu ^+ \mu ^-} > 120$\,GeV} & 29 (72{\%}) & 0.1 & 600 & 4.4  \\
\hline
\end{tabular}
\label{tab:DetailsCutFlow}
\end{table}


Finally, it should be noted that in the $\nu \overline{\nu} H$ process, especially at $\sqrt{s} =500$\,GeV, two signal processes ($ZH$ process with $Z \to \nu \overline{\nu}$ and $WW$-fusion process) are contributing.
The relative contributions of these production modes will be fixed to the percent-level or better from other Higgs decay modes like $H \to b \overline{b}$ and can be used to convert the cross section times branching fraction measurement into a measurement of $\mathrm{BF}(H \to \mu^+\mu^-)$.
With the help of the total $ZH$ cross section determined with the recoil method, the absolute $H \mu \mu$ Yukawa coupling can be extracted.
Therefore, the quoted ILC precisions can directly be taken as precision on the branching fraction for $H \to \mu^+\mu^-$, or, divided by a factor of two, as precision on the muon Yukawa coupling.
This is qualitatively different from the signal strength determinations at the (HL-)LHC.

\section{Impact of the Transverse Momentum Resolution}
\label{sec:impact:ptres}
The di-muon mass $M_{\mu ^+ \mu ^-}$ is the most important observable for this analysis.
The uncertainty on $M_{\mu ^+ \mu ^-}$ is directly related to the precision of the measurement of the muon momentum, and in particular the resolution on its transverse component, $\sigma _{1/P_t}$, plays a crucial role in this analysis.
The transverse momentum resolution of the ILD detector has been shown already in Fig.~\ref{fig:DBD_momres} as a function of the momentum for different polar angles.

Instead of implementing the full $p$ and $\theta$ dependency of the resolution, a simplified approach of smearing all true muon transverse momenta with the same resolution has been taken here.
This is a fully justified approach in case of $\sqrt{s} =500$\,GeV, since the vast majority of muons have high momenta in the asymptotic regime, and, due to the isotropic decays of the Higgs boson, are mostly at large polar angles in the centre of the detector.
For the case of $\sqrt{s} =250$\,GeV, the muons have lower momenta between $40$ and $100$\,GeV and the approximation is less precise, but still useful.

The dependence of the result on the asymptotic value of the transverse momentum resolution has been studied by adding a Gaussian-distributed error to the transverse momentum taken from the MC-truth information for all events passing the preselection described in Sec.~\ref{subsec:analysis:presel}.
All other quantities in the event are taken from the full simulation as before.
Transverse momentum resolutions between $1 \times 10^{-3}$ to $1 \times 10^{-6}$ (GeV$^{-1}$) have been considered.
The background is kept unchanged from the full simulation study since its invariant mass distribution after the BDTG score cut does not exhibit any sharp peaks, as can be seen, \textit{e.g.}: in Fig.~\ref{fig:CBGfit_nnh500L}, and thus a change in momentum resolution will not affect the distribution significantly.

\begin{figure}[t]
\centering
\includegraphics[width = 13truecm]{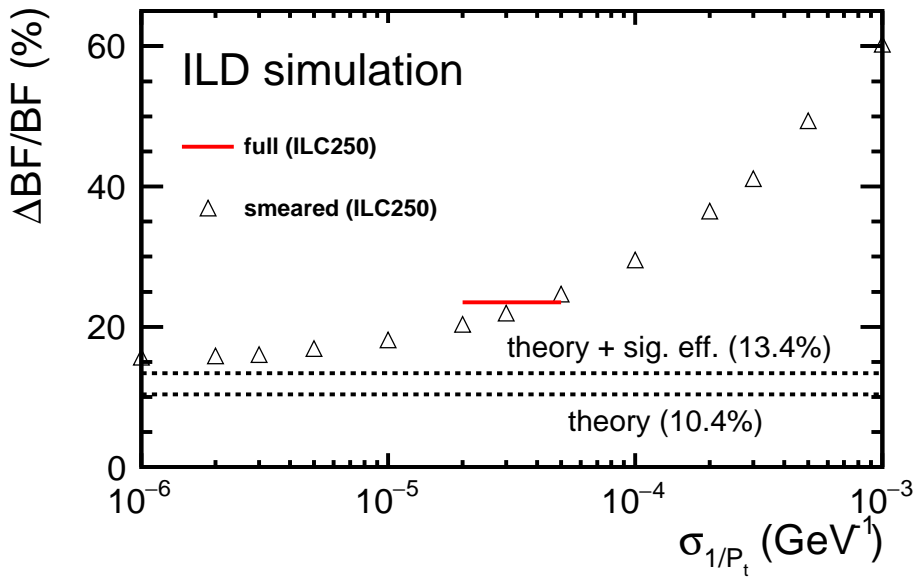}
\caption{Expected precision on $\mathrm{BF}(H \to \mu ^+ \mu ^-)$ as a function of transverse momentum resolution $\sigma _{1/P_t}$ (triangles), together with full simulation results discussed in Sec.~\ref{subsec:analysis:result} (red line) and the theoretical limits defined in Sec.~\ref{subsec:analysis:result} (dashed lines) for 2~ab$^{-1}$ collected at $\sqrt{s} = 250$\,GeV.
The red line indicates the typical transverse momentum resolution range at $\sqrt{s} = 250$\,GeV.}
\label{fig:momresimpact_250}
\end{figure}

Fig.~\ref{fig:momresimpact_250} shows the obtained precision on $\mathrm{BF}(H \to \mu ^+ \mu ^-)$ as a function of the transverse momentum resolution $\sigma_{1/P_t}$ at $\sqrt{s} = 250$\,GeV, together with the theoretical limit as defined in Sec.~\ref{subsec:analysis:result} shown by dashed lines.
The red line indicates the typical value of transverse momentum resolution at $\sqrt{s} = 250$\,GeV.
The ``effective'' resolution for which the smearing approach gives the same precision on the branching fraction as the full simulation is $\sim 4 \times 10^{-5}$.
This result is consistent with Fig.~\ref{fig:DBD_momres}, because at this energy muons typically have momenta in the regime of 40 to 100\,GeV which corresponds to a resolution of around $\sim 4 \times 10^{-5}$.

The following conclusion can be drawn.
First of all, with $\sigma _{1/P_t} = 2 \times 10^{-4}$\,GeV$^{-1}$ for example, typical for LHC experiments, precision would be 36{\%} instead of 23{\%}, \textit{i.e.} bigger by a factor of 1.6.
Therefore, it is very important for this analysis to reach the ILD goal for the transverse momentum resolution.
In the other direction, though technologically not realistic, an improvement of $\sigma_{1/P_t}$ to a few times $10^{-6}$\,GeV$^{-1}$ would allow to nearly reach the ``zero-background'' scenario, in the sense that although the same amount of (irreducible) background events pass the selection, the Higgs signal peak becomes so narrow that the background contribution underneath the peak doesn't have a significant effect anymore.


\begin{figure}[t]
\centering
\includegraphics[width = 13truecm]{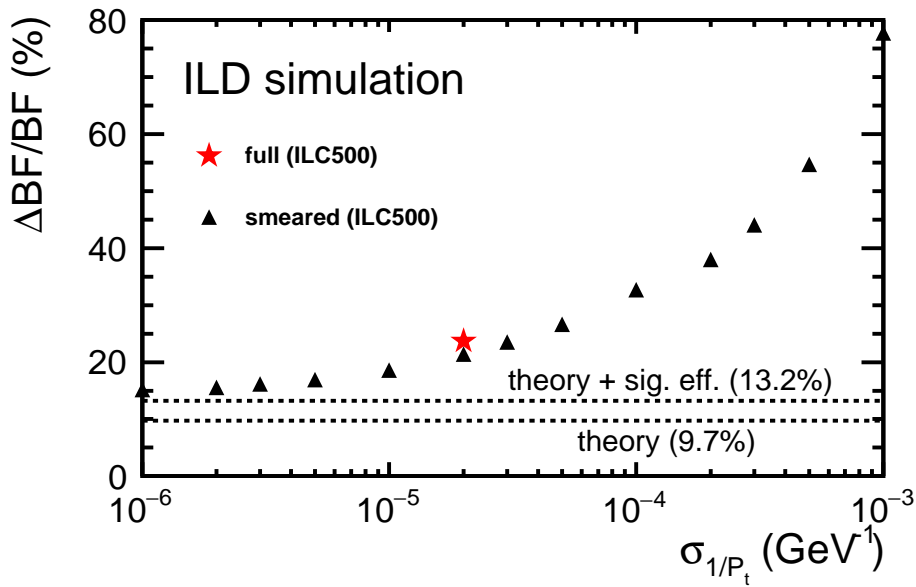}
\caption{Expected precision on $\mathrm{BF}(H \to \mu ^+ \mu ^-)$ as a function of transverse momentum resolution $\sigma _{1/P_t}$ (triangles), together with full simulation results discussed in Sec.~\ref{subsec:analysis:result} (star) and the theoretical limits defined in Sec.~\ref{subsec:analysis:result} (dashed lines) for 4~ab$^{-1}$ collected at $\sqrt{s} = 500$\,GeV.}
\label{fig:momresimpact_500}
\end{figure}

\begin{figure}[t]
\centering
\includegraphics[width = 13truecm]{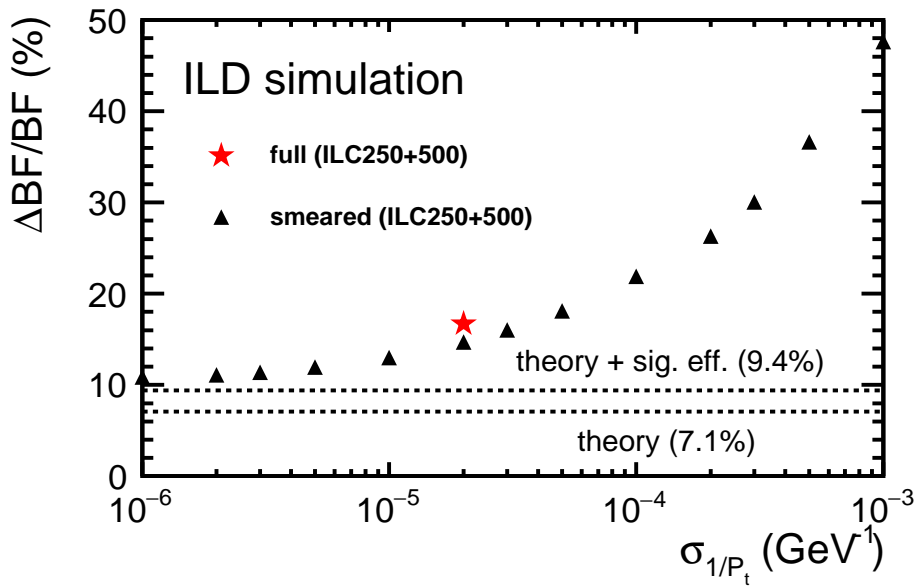}
\caption{Expected precision on $\mathrm{BF}(H \to \mu ^+ \mu ^-)$ as a function of transverse momentum resolution $\sigma _{1/P_t}$ (triangles), together with full simulation results discussed in Sec.~\ref{subsec:analysis:result} (star) and the theoretical limits defined in Sec.~\ref{subsec:analysis:result} (dashed lines) for the combination of the 2~ab$^{-1}$ collected at $\sqrt{s} =250$\,GeV and 4~ab$^{-1}$ collected at $\sqrt{s} = 500$\,GeV data sets.}
\label{fig:momresimpact}
\end{figure}

Fig.~\ref{fig:momresimpact_500} shows the equivalent result at $\sqrt{s} =500$\,GeV, while the combined result of both centre-of-mass energies is displayed in Fig.~\ref{fig:momresimpact}. Since at $\sqrt{s} =500$\,GeV the momenta of the muons are higher than in the $\sqrt{s} = 250$\,GeV case, the transverse momentum resolution is closer to the asymptotic performance of $2 \times 10^{-5}$,  and thus the ``effective'' resolution gets closer to the case of $2 \times 10^{-5}$. Otherwise, the conclusions remain similar to the $\sqrt{s} = 250$\,GeV case, underlining again the importance to achieve the ILD design goal on the transverse momentum resolution.


A similar study has also been performed by the Compact LInear Collider (CLIC)~\cite{mu4}, based on $e^+ e^- \to \nu \overline{\nu} H$ at $\sqrt{s} = 1.4$\,TeV (the $q\overline{q}h$ contribution is negligible at $1.4$\,TeV).
Figure~11 in Ref.~\cite{mu4} shows a saturation around $\sigma _{1/P_t} = 1 \times 10^{-5}$, where the precision on $\mathrm{BF}(H \to \mu ^+ \mu ^-)$ reaches $\sim 25{\%}$.
While the saturation is reached already for worse $\sigma _{1/P_t}$ resolutions compared to the ILD case, the CLIC study leads to the same conclusion as our analysis, namely that even a large improvement of the muon momentum resolution would result in only a moderate improvement of the statistical uncertainty of the measured product of the Higgs production cross-section and the branching fraction for the $H \to \mu ^+ \mu ^-$ decay.
On the other hand, not reaching the design goal for the momentum resolution would lead to a significant loss of sensitivity. 


\section{Summary}
\label{sec:summary}
In this study, the prospects for measuring the branching fraction of $H \to \mu ^+ \mu ^-$ at the ILC have been evaluated based on full simulation of the ILD detector for the $\sqrt{s} =250$\,GeV and 500\,GeV data sets as defined by the standard ILC running scenario.
Eight channels have been analysed in total, $\sqrt{s}$ of 250\,GeV and 500\,GeV, two beam polarisation cases, and the two signal processes $q\overline{q}H$ and $\nu \overline{\nu} H$.
The combined precision on $\mathrm{BF}(H \to \mu ^+ \mu ^-)$ using all results is estimated to be 17{\%}; the 250\,GeV data alone yield a precision of 23{\%}.
These results are about a factor of 2.4 bigger than the ``theoretical'' limit of zero background and $100\%$ efficiency. 
Compared to most recent HL-LHC prospects based on the full detector upgrades, the precision is only about $50\%$ larger, despite the fact that $100$ times more $H \to \mu^+ \mu^-$ events are expected to be produced at HL-LHC.  In combination with other ILC measurements, the observed signal strength can be translated into a direct measurement of the branching fraction and thus the muon Yukawa coupling.
The combination of HL-LHC and ILC full program up to $1$\,TeV would provide an ultimate precision of $\sim 7{\%}$ on BF($H \to \mu^+ \mu^-$).
In addition to the full simulation analysis, the impact of the transverse momentum resolution was studied.
This study shows the importance to achieve the ILD design goal of the transverse momentum resolution, otherwise the precision will be significantly degraded.
The first evaluation of the prospects to measure $\mathrm{BF}(H \to \mu ^+ \mu ^-)$ at the lower-energy stages of the ILC presented in this paper could be improved in future analyses. Interesting points to study comprise the application of beam-spot constraint in the track fit of the two muons, a full treatment of events with significant FSR and the inclusion of the $Z$ boson decays to charged leptons.

\section*{Acknowledgements}
We would like to thank the LCC generator working group and the ILD software working group for providing the simulation and reconstruction tools and producing the Monte Carlo samples used in this study.
SK would like to thank Junping Tian (The University of Tokyo) for lots of useful comments and discussing technical details in the analysis.
This work has benefited from computing services provided by the ILC Virtual Organization, supported by the national resource providers of the EGI Federation and the Open Science GRID.
We thankfully acknowledge the support by the 
the Deutsche Forschungsgemeinschaft (DFG, German Research Foundation) under Germany's Excellence Strategy EXC 2121 ``Quantum Universe'' 390833306.

\printbibliography{}
\end{document}